\documentclass[useAMS,usenatbib]{mn2e}
\usepackage{graphicx}
\newcommand\Msun{M$_\odot$}
\newcommand\Mstar{M$_\star$}
\newcommand\ha{H\,{$\alpha$}}
\newcommand\hb{H\,{$\beta$}}
\newcommand\oiii{[O\,{\sevensize III}]}
\newcommand\oii{[O\,{\sevensize II}]}
\newcommand\nii{[N\,{\sevensize II}]}
\newcommand\sigfifth{$\Sigma_{\rm 5th}$}
\newcommand\Aha{$A_{\rm H\alpha}$}

\title[XMMXCS J2215.9-1738 cluster]{
Properties of star-forming galaxies in a cluster and its surrounding structure at $\bmath{z=1.46}$
}
\author[M. Hayashi et al.]{%
Masao Hayashi,$^{1}$\thanks{E-mail: masao.hayashi@nao.ac.jp}
Tadayuki Kodama,$^{1,2}$
Yusei Koyama,$^{1,3}$
\newauthor
Ken-ichi Tadaki,$^{1,3}$
Ichi Tanaka$^{2}$\\
$^{1}$Optical and Infrared Astronomy Division, National Astronomical Observatory, Mitaka, Tokyo 181-8588, Japan\\
$^{2}$Subaru Telescope, National Astronomical Observatory of Japan, 650 North A'ohoku Place, Hilo, HI 96720, USA\\
$^{3}$Department of Astronomy, Graduate School of Science, University of Tokyo, Tokyo 113-0033, Japan
}

\begin{document}

\date{Accepted 2011 April 8. Received 2011 April 8; in original form 2011 February 15}

\pagerange{\pageref{firstpage}--\pageref{lastpage}} \pubyear{2011}

\maketitle

\label{firstpage}

\begin{abstract}
We conduct a wide-field narrow-band imaging survey of \oii\ emitters
in and around the XMMXCS J2215.9-1738 cluster at $z=1.46$ with
Subaru/Suprime-Cam. In a 32\arcmin$\times$23\arcmin\ area, we select
380 \oii\ emitting galaxies down to $1.4\times10^{-17}$ erg s$^{-1}$
cm$^{-2}$. Among them, 16 \oii\ emitters in the central region of the
cluster are confirmed by near-infrared spectroscopy with
Subaru/MOIRCS, suggesting that our photometric selection is valid to
sample \oii\ emitters at $z=1.46$.
We find that \oii\ emitters are distributed along filamentary
large-scale structures around the cluster, which are among the largest
structures of star-forming galaxies ever identified at $1.3 \la z \la 3.0$. 
We define several environments such as cluster core, outskirts,
filament, and field, in order to investigate the environmental
dependence of star-forming galaxies at $z=1.46$.       
The colour--magnitude diagram of $z'-K$ vs.~$K$ for the \oii\ emitting
galaxies shows that a significantly higher fraction of \oii\ emitters
with red $z'-K$ colours is seen in the cluster core than in other
environments. It seems that the environment which hosts such red
star-forming galaxies shifts from the core region at $z=1.46$ to the
outskirts of clusters at lower redshifts. 
The multi-colour analysis of the red emitters indicates that these
galaxies are more like nearly passively evolving galaxies which host
\oii\ emitting AGNs, rather than dust-reddened star-forming \oii\
emitters. We argue therefore that AGN feedback may be one of the
critical processes to quench star formation in massive galaxies in
high density regions. 
The emission line ratios of \oiii/\hb\ and \nii/\ha\ of the \oii\
emitters in the cluster core support that there is a moderate contribution
of AGN to the emitters. We also find that the cluster has experienced
high star formation activities at rates comparable to that in the
field at $z=1.46$ in contrast to lower redshift clusters, and that
star formation activity in galaxy clusters on average increases with
redshift up to $z=1.46$.        
In addition, line ratios of \nii/\ha\ and \oiii/\hb\ indicate that a
mass--metallicity relation exists in the cluster at $z=1.46$, which is
similar to that of star-forming galaxies in the field at
$z\sim2$. These results all suggest that at $z\sim1.5$ star formation
activity in the cluster core becomes as high as those in low density
environments and there is apparently not yet a strong environmental
dependence, except for the red emitters.
\end{abstract}


\section{Introduction}

Recently, galaxy clusters at high redshift of $z>1$, especially
$z\ga1.5$, are found one after another 
\citep{kurk2009,papovich2010,tanaka2010,henry2010,gobat2011,fassbender2011}.     
These high-$z$ clusters provide important clues to understanding
galaxy formation and evolution, since the redshift of $z>1$ approaches
the epoch when galaxy clusters were formed. 
The importance of $z\ga1$ on galaxy evolution is also supported by
the fact that the cosmic star formation activity and the number
density of active galactic nuclei (AGN) both come to a peak at $z$=1--3
\citep[e.g.,][]{ueda2003,hopkins2006}, and this redshift range
corresponds to the epoch when galaxies and AGNs are evolving vigorously.
In galaxy clusters at low and intermediate redshifts of $z<1$,
a prominent sequence of red galaxies is seen in a color-magnitude diagram
\citep[e.g.,][]{bower1992,kodama1998,stanford1998,vandokkum1998,blakeslee2003,tanaka2005,deLucia2007}. 
It is also well-known that clusters are dominated by early-type galaxies.
On the other hand, in proto-clusters at high redshift of $z>3$,
overdense regions of star-forming galaxies
such as Lyman alpha emitters and Lyman break galaxies are reported
\citep[e.g.,][]{steidel1998}. \citet{kodama2007} found a deficit of
massive galaxies on red sequence in proto-clusters at $z\sim3$, while
such red galaxies are already in place in proto-clusters at $z\sim2$
\citep[see also][]{kajisawa2006,kriek2008,doherty2010}. 
These results suggest that the blue star-forming galaxies in the early
phase evolve to red passive galaxies in high density regions
during the redshift interval of 1--3.
In fact, recent studies show that clusters at $z\ga1.5$
are conducting active star formation even in the core regions
\citep{hayashi2010,hilton2010,fassbender2011,papovich2010,tran2010}.
It is thus important to reveal star formation and AGN activities in
galaxy clusters at this epoch.  

An effective method of selecting star-forming galaxies is to search for
emission line galaxies based on narrow-band imaging, which enables us to
sample star-forming galaxies with line emission stronger than a certain
limiting flux and an equivalent width. A wide-field imaging is also essential
to cover cluster outskirts and surrounding field regions as well as 
cluster cores for environmental studies.
Recent studies have discovered that the medium-density regions in
the outskirts of galaxy clusters are the key environment to determine
galaxy properties such as colours and star formation activities during the course
of cluster assembly \citep{tanaka2005,koyama2008}. They found that the colours of
galaxies change sharply from blue to red in such medium-density
regions. Furthermore, \citet{koyama2008} found that it is possible that
star formation is the most active in the medium-density regions at
$z=0.81$.  These studies have demonstrated the importance of wide-field
surveys comprehensively covering all the range in environments from 
cluster cores to the surrounding fields.

Moreover, follow-up spectroscopy of the emitters detected in the
narrow-band imaging is essential to characterize their detailed properties
such as dust-corrected star formation rate (SFR), AGN contribution and
gaseous metallicity.
At $z=1-3$, because all the useful emission lines such as \ha, \hb,
and \oiii\ are redshifted into the near-infrared regime,
deep multi-object near-infrared (NIR) spectroscopy becomes critically important
and effective.
Dust correction is one of the major uncertainties
in characterizing galaxy properties from the observed data, and 
the Balmer decrement measurement (\hb/\ha) with NIR spectroscopy is essential
to make accurate correction for dust extinction.
Moreover, recent NIR spectroscopic observations of star-forming galaxies at 
high redshifts have revealed that there is a mass--metallicity relation
up to $z\sim3$, and that the chemical evolution is seen in the sense that
galaxies with a given stellar mass have lower metallicities on average
with increasing redshift \citep{maiolino2008,mannucci2009}. 
More recently, \citet{mannucci2010} discovered a fundamental relationship
between stellar mass, metallicity, and SFR, which indicates that
galaxies with smaller stellar masses and/or higher SFRs have lower metallicities.
It should be noted that metallicities of galaxies also depend on the
sample selection \citep[e.g.,][]{hayashi2009,yoshikawa2010,onodera2010}.  
All these previous studies were, however, conducted in general fields,
and it is yet unknown whether such mass--metallicity relation is
established in clusters and how it depends on environment.
Metallicity of star-forming galaxies provides information
on the integrated star formation history in galaxies and it is independent
from the snap-shot measurement of on-going star formation rate at each epoch.
Therefore, the NIR spectroscopy of star-forming galaxies in various environments
at $z>1$ is the key to understanding galaxy evolution and its environmental
dependence in much greater detail.

We have been conducting deep and wide-field surveys of \ha\ and \oii\
emitters in some general fields and in galaxy clusters at various redshifts 
at $z>0.4$ as MAHALO-Subaru project (MApping HAlpha and Lines of Oxygen with
Subaru; PI is T. Kodama). Among the clusters targeted by our project,
XMMXCS~J2215.9-1738 cluster (hereafter XCS2215 cluster) at $z=1.46$ 
\citep{stanford2006} is one of the most massive galaxy clusters at high
redshift of $z>1$. \citet{hayashi2010} reported a deep survey
of \oii\ emitters in the central 6\arcmin$\times$6\arcmin\ region of
the XCS2215 cluster, and found that there are a lot of star
forming galaxies even in the cluster core where almost no star formation
is seen in lower-$z$ clusters.  
\citet{hilton2010} found eight mid-infrared (24\micron)
sources in the central region of the XCS2215 cluster with Spitzer/MIPS,
which supports our results that a relatively large fraction of galaxies
in the XCS2215 cluster are still having active star formation. 
The previous \oii\ survey \citep{hayashi2010} was limited to the central
6$'$$\times$6$'$ area where we had the MOIRCS $K_s$-band data as well.
In this paper, we expand our previous survey to the outskirts of this cluster
in order to investigate the environmental dependence of galaxy
properties at $z=1.46$.
Our new UKIRT wide-field $K$-band imaging data now enable us to
perform the analysis over the full Suprime-Cam field across
$\sim$30~arcmin in diameter.
At the same time we perform follow-up NIR
spectroscopy of \oii\ emitters identified in the central region of the
cluster to examine their properties in detail.

The structure of this paper is as follows.
Observations and available data for the XCS2215 cluster are described
in \S~\ref{sec;data}. Then, \oii\ emitters at $z=1.46$ around the
cluster are selected from the photometric catalogues in
\S~\ref{sec;select-oii}. Some of the \oii\ emitters are confirmed by
spectroscopy in \S~\ref{sec;spec-confirm}. 
In \S~\ref{sec;phot-properties}, environmental dependence of colour,
SFR, and specific SFR are discussed, and we compare the star forming
activity of this cluster to those of lower-$z$ clusters in
\S~\ref{sec;evolution_sf}.  
In \S~\ref{sec;spec-properties}, we investigate AGN contribution, dust
extinction, and gas-phase metallicity based on the emission line
ratios between \ha, \hb, \oiii, and \nii\ lines. Finally, in
\S~\ref{sec;conclusions}, we summarize our results of this paper and
make conclusions. 
Throughout this paper, magnitudes are presented in the AB system, and we adopt
cosmological parameters of $h=0.7$, $\Omega_{m}=0.3$ and
$\Omega_{\Lambda}=0.7$. Vega magnitudes in $J$ and $K$, if
preferred, can be obtained from our AB magnitudes using the relations:
$J$(Vega)=$J$(AB)$-$0.92 and $K$(Vega)=$K$(AB)$-$1.90, respectively.
In $z=1.46$, 1 arcmin corresponds to 1.25 Mpc (comoving) and 0.51 Mpc
(physical), respectively. 

\begin{table*}
 \caption{Summary of the optical and NIR imaging data. The limiting
 magnitudes are measured with a 2\arcsec diameter aperture. 
 Note that $J$-band data are available only in the central region of
 the cluster, and its depth depends on the position due to non-uniform
 integration times. The PSFs in all the co-added images are 
 matched to 1.09 arcsec.
  }
\begin{center}
\begin{tabular}{cccccc}
\hline\hline
filter & effective area & net integration & limiting mag. & instrument & observation date\\
       & (arcmin$^2$)  & (minutes)       & (3$\sigma$)   & &   \\
\hline
$B$    & 32$\times$23  & 140 & 27.6 & Subaru/Suprime-Cam & 2008/07/30-31$^\dagger$\\
$R_c$  & 32$\times$23  & 88  & 27.1 & Subaru/Suprime-Cam & 2009/07/20 \\
$i'$   & 32$\times$23  & 90  & 26.8 & Subaru/Suprime-Cam & 2009/07/20 \\
$z'$   & 32$\times$23  & 80  & 25.8 & Subaru/Suprime-Cam & 2008/07/30-31$^\dagger$ \\
NB912  & 32$\times$23  & 260 & 25.8 & Subaru/Suprime-Cam & 2008/07/30-31$^\dagger$ \\
$J$    &  6$\times$6   & 32.5 -- 92.3 & 23.8 -- 24.6 & Subaru/MOIRCS &
 2007/08/07$^\ddagger$, 2008/06/29-30$^\dagger$\\
$K$    & 32$\times$23  & 123 & 23.4 & UKIRT/WFCAM  & 2010/07/30-31\\
\hline\hline
\multicolumn{6}{l}{$\dagger$ \citet{hayashi2010}, $\ddagger$ \citet{hilton2009}}
\label{tbl;imaging-data}
\end{tabular}
\end{center}
\end{table*}

\section{Observations and Data}
\label{sec;data}

\subsection{Optical imaging data}

We have obtained optical imaging data of the XCS2215 cluster with
Subaru Prime Focus Camera \citep[Suprime-Cam;][]{miyazaki2002} on the
Subaru telescope. Suprime-Cam consists of ten 2048$\times$4096 CCDs
with a pixel scale of 0.20\arcsec, and has a wide field-of-view (FoV)
of 34$\times$27 arcmin$^2$, which enables us to cover from the central
region to the outskirts of the cluster by a single pointing.
Our multi-wavelength data set consists of four
broad-band images in $B$, $R_c$, $i'$, $z'$ bands, and a narrow-band
image of NB912 ($\lambda_c$=9139\AA, $\Delta\lambda$=134\AA).  

The observations were conducted in two semesters. The $B$, $z'$, and
NB912-band imaging were performed under a normal open-use program (S08A-011,
PI: T. Kodama) on 2008 July 30-31, while $R_c$ and $i'$-band imaging
were performed under a service-mode open-use program (S09A-168S, PI: M. Hayashi) 
on 2009 July 20. The details of the observation and data reduction for
$B$, $z'$ and NB912 data are described in \citet{hayashi2010}. In this paper,
therefore, we mainly describe the additional data in $R_c$ and
$i'$. The individual exposure times of a frame in $R_c$ and $i'$ were
8 and 6 minutes, and the total integration times were 88 and 90
minutes, respectively. The weather was fine during the
observations. The sky condition was mostly photometric, although thin
cirrus occasionally passed the targeted location.
The seeing was 0.6--0.8 arcsec in both $R_c$ and $i'$.   

The data reduction of $R_c$ and $i'$ was carried out in the same
manner as for the $B$, $z'$, and NB912 data, and we used the data
reduction package for Suprime-Cam 
\citep[{\sc sdfred} ver.1.4:][]{yagi2002,ouchi2004}.  
The point spread functions (PSFs) in the $R_c$ and $i'$ images
were matched to
1.09 arcsec, which was the PSF size of the other optical data.   
The photometric zero-points were determined using the
photometric standard stars in SA113 \citep{landolt1992}. 
The 3$\sigma$ limiting magnitudes were 27.1 and 26.8 in $R_c$ and $i'$,
respectively. The specifications of all the optical data are summarized in
Table \ref{tbl;imaging-data}.

\subsection{Near-infrared imaging data}

\setcounter{footnote}{3}

The NIR imaging data are updated from
\citet{hayashi2010}, where we had $J$ and $K_s$-band data covering
only the central 6\arcmin$\times$6\arcmin\ region with Multi-Object
Infrared Camera and Spectrograph
\citep[MOIRCS;][]{ichikawa2006,suzuki2008} on the Subaru telescope. 
MOIRCS consists of two 2048$\times$2048 HgCdTe detectors with a pixel
scale of 0.117\arcsec, and its FoV is 4\arcmin$\times$7\arcmin.

We have now obtained a new wide-field $K$-band data with
Wide Field Camera \citep[WFCAM;][]{casali2007} on the United Kingdom
Infrared Telescope (UKIRT) on 2010 July 30--31 (U/10A/J3, PI:
Y. Koyama) in order to cover the entire region of the optical data. WFCAM
consists of four 2048$\times$2048 HgCdTe detectors with a pixel scale
of 0.4\arcsec, and each detector can cover
13.65\arcmin$\times$13.65\arcmin\ region. Because the detectors are
spaced with a gap of 12.83\arcmin, four pointings are required to
get a contiguous sky coverage of a 0.75 square degrees. A dither pattern of
five points with 2$\times$2 small microstepping was set for each
pointing, and so the individual exposures were conducted at 20 different
positions for a cycle at each pointing. The exposure time of each frame was
10 seconds, and the total integration time at each pointing was 123 minutes.
The weather was fine throughout the two days of the observing run,
and the sky condition was photometric. The seeing was $\sim$1.0 arcsec.      

We reduced the WFCAM data in a standard manner
using our own IRAF-based software (provided by K. Motohara). 
First, we subtracted a dark frame from individual object frames, and
then a self-flat image was created by combining 80 frames
(4 cycles) and taking a median value at each pixel.
After flat-fielding, all the object frames were
temporarily mosaiced and co-added. We then conducted a source
detection on the co-added image. Next, after masking the detected objects
in the individual frames, we make self-flat images again,
and then conduct a flat-fielding per each observation cycle of 20 frames.   
After the sky background was subtracted from each frame, 20 frames of
each cycle are combined. Finally, PSF sizes of the images of each cycle 
are matched, and the frames were mosaiced and co-added to make the final
images. A special care was taken to exclude spurious objects due to
crosstalk of bright objects. PSF in $K$ was matched to 1.09 arcsec. The
photometric zero-point was determined with 2MASS catalog
\citep{skrutskie2006}. The 3$\sigma$ limiting magnitude was 23.3 in
$K$. The specifications of the NIR data are also summarized in Table 
\ref{tbl;imaging-data}     

For $J$-band, we added the archival data with MOIRCS by
\citet{hilton2009} to our own data. \citet{hilton2009} observed a
4\arcmin$\times$4\arcmin\ region of the cluster core with the chip 2
only of MOIRCS with an exposure time of 1485 seconds in total.
Combining the archive data with our own data,
we conducted the data reduction again in the same manner as
in \citet{hayashi2010} using the data reduction package for
MOIRCS ({\sc mcsred}\footnote{http://www.naoj.org/staff/ichi/MCSRED/mcsred.html} 
by I.~Tanaka et~al.).
PSF in $J$ was matched to 1.09 arcsec. 

\begin{table}
 \caption{Summary of the spectroscopic observations with Subaru/MOIRCS.}
\begin{center}
\begin{tabular}{ccccc}
\hline\hline
slitmask & \# of target & grism & integration & seeing \\
         &  & & (minutes) & (arcsec) \\
\hline
MOS1 & 17 & $zJ500$ & 180 & 0.55-0.80 \\
MOS2 & 17 & $zJ500$ & 300 & 0.45-0.60 \\
\hline\hline
\label{tbl;spec-data}
\end{tabular}
\end{center}
\end{table}

\subsection{Near-infrared spectroscopic data}

We conducted a follow-up NIR spectroscopy of \oii\ emitters in the
central region of the cluster with MOIRCS on the Subaru telescope
on 2009 September 3--7 (S09B-012, PI: T. Kodama). For 34 out of 44
\oii\ emitter candidates identified in \citet{hayashi2010}, we obtained NIR
low-resolution spectra with $zJ500$ grism 
with a resolution of $R=700$ at $J$-band and a wavelength coverage
of 0.9--1.8\micron. The dispersion was 5.57 \AA~pixel$^{-1}$. 
We used two masks, and 17 slits were allocated to the \oii\ emitter
candidates in each mask.
The width and the length of individual slits were 0.7\arcsec
and 11-12\arcsec, respectively.  Depending on the positions of slits on
the mask, 24 spectra have a wavelength coverage of 1.0--1.7\micron, which
can neatly cover all the redshifted \ha, \hb, \oiii, and \nii\ emission lines
if present.  The remaining 10 spectra have a coverage of
1.0--1.35\micron, which covers \hb, and \oiii\ emission lines.
Moreover, about two-thirds of the spectra, i.e.\ 21 \oii\ emitters,
extend down to 0.9\micron. The specifications of the
spectroscopic observations are summarized in Table~\ref{tbl;spec-data}.

We observed the targets consecutively at two positions (A and B) with
a offset of 3\arcsec\ on the slit. Single exposure time was 600 or 900
seconds depending on the sky background level. The total on-source
integration time was 3 hours for the MOS1 mask and 5 hours for the
MOS2 mask. We also observed a standard star BD+17\degr4708 \citep{bohlin2004}
on each night to correct for a telluric absorption as well as the instrumental
efficiency, and to make flux calibration. Most of the time during the observations,
however, the weather condition was not good and the sky was
covered with thin cirrus and not photometric.
Therefore, the absolute flux calibration was very difficult for these spectra.
The seeing was 0.45--0.80\arcsec in FWHM.

The data reduction was done with standard procedures using {\sc iraf}. First,
bad pixels and cosmic rays were removed from each frame, and 
a A$-$B frame was created from a pair of successive frames observed
at the two positions A and B. Then, flat-fielding was done with a dome-flat
image for the individual A$-$B images.
Next, distortion was corrected using a calibration data provided by
the MOIRCS instrument team. 
After extracting a spectrum at each slit, wavelength
calibration was done with the OH airglow lines. Then, residual sky
subtraction was carried out, since the A$-$B procedure alone might not
completely remove the sky background due to its time variation. All
the spectra for each \oii\ emitter were then co-added and an one-dimensional
spectrum was extracted by combining ten pixels along a slit.
Finally, the telluric absorption and the instrumental efficiency were
corrected using the spectra of BD+17\degr4708.  
As an error, the sky noise was estimated as a square root of the photon
count on the sky spectrum. 

\subsection{Photometric catalogue}

\begin{figure}
 \begin{center}
 \includegraphics[width=\linewidth]{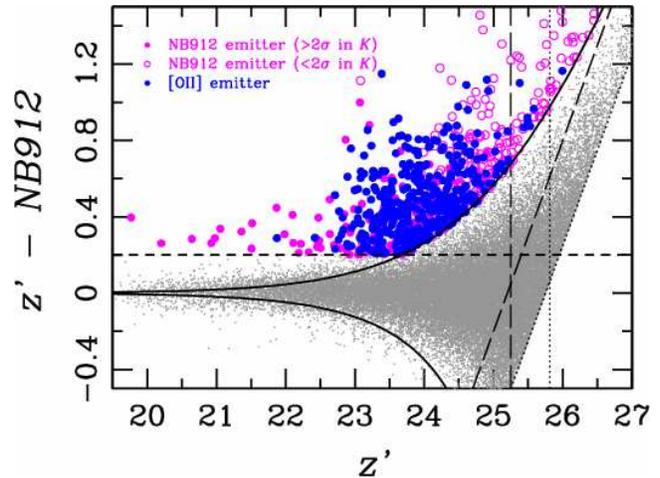}
 \end{center}
 \caption{colour--magnitude diagram of $z'$ vs. $z'-NB912$.
 Solid lines show 3$\sigma$ excesses of $NB912$ over $z'$. Long-dashed
 and dotted lines show 5$\sigma$ and 3$\sigma$ limiting magnitudes, 
 respectively. Short-dashed line shows $z'-NB912=0.2$. Gray dots show
 galaxies brighter than 5$\sigma$ in NB912. Magenta filled circles indicate
 NB912 emitters with more than 2$\sigma$ detections in $K$, while magenta
 open circles indicate those without detection in $K$. Blue filled circles
 show our 380 \oii\ emitters in the XCS2215 cluster.}  
 \label{fig;zznb}
\end{figure}

\begin{figure*}
 \includegraphics[width=0.495\linewidth]{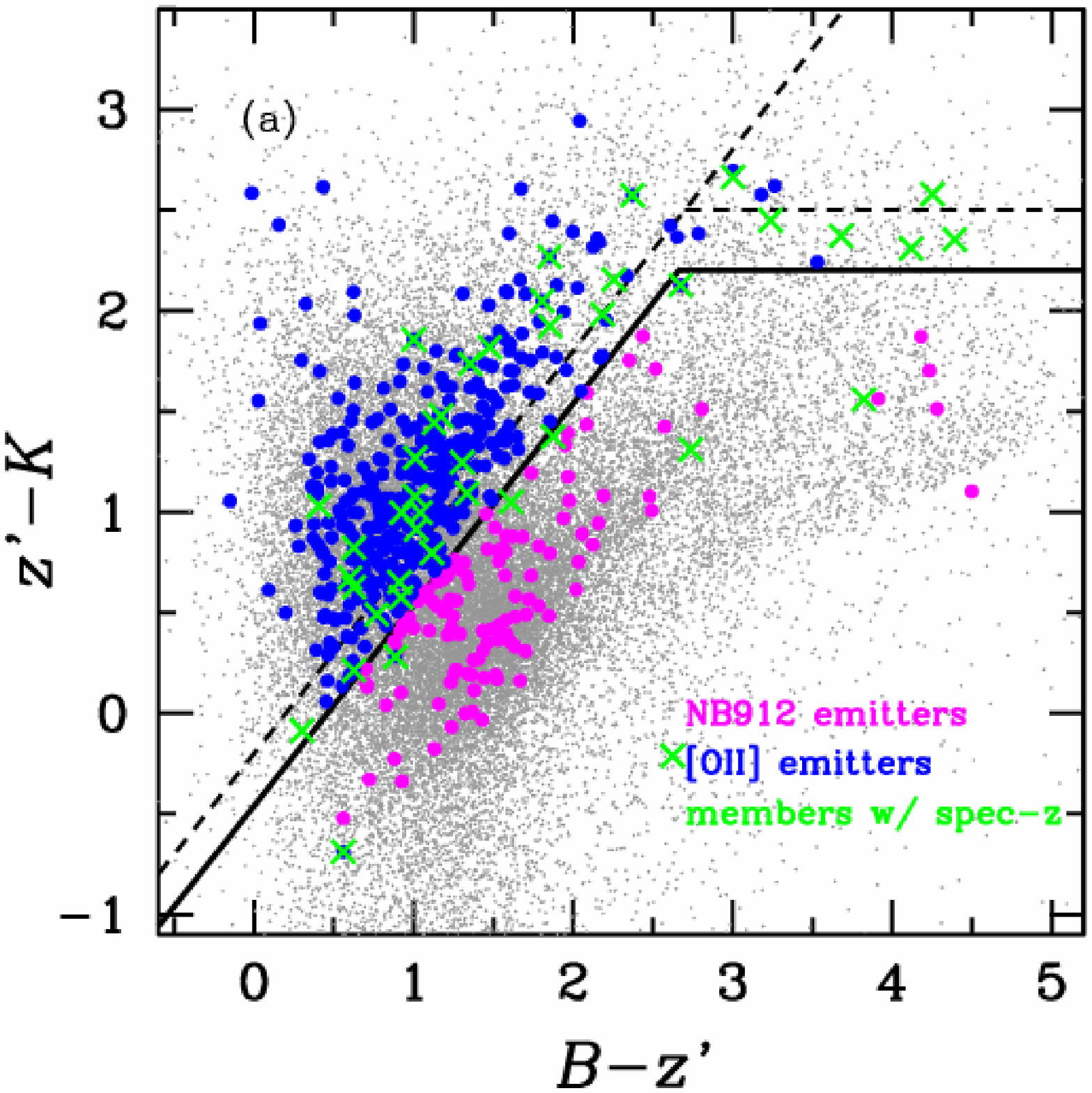}
 \includegraphics[width=0.495\linewidth]{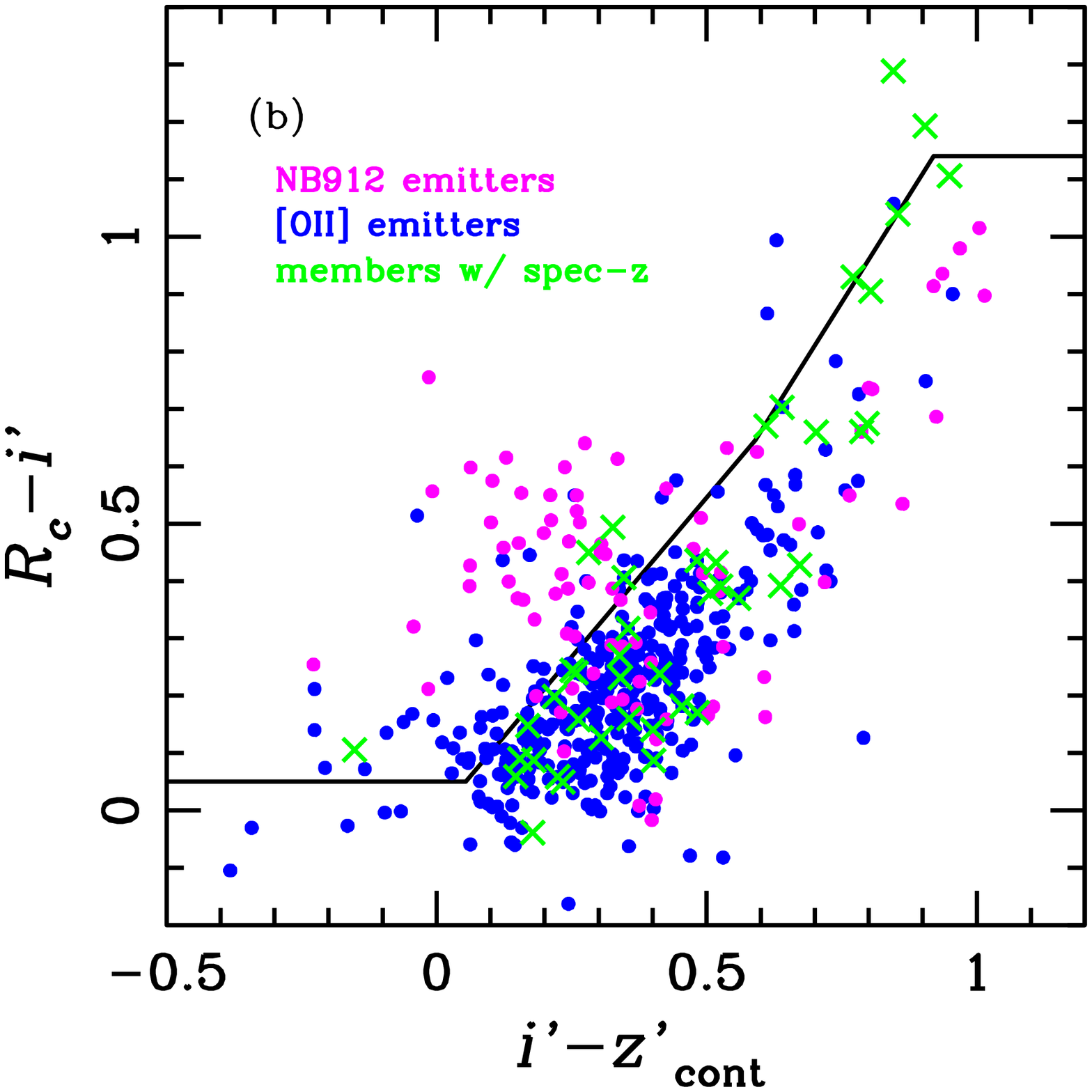}
 \caption{(a) Left panel: the colour-colour diagram of $B-z'$
   vs. $z'-K$.  The solid lines show our criteria for the selection of
   \oii\ emitters at $z\sim1.46$, while the broken lines indicate the
   original $BzK$ selection criteria defined by
   \citet{daddi2004}. Blue filled circles show 380 \oii\ emitters, and 
   magenta filled circles show NB912 emitters other than \oii\ emitters. Gray
   dots are galaxies brighter than 5$\sigma$ detection in NB912, and green
   crosses indicate cluster members confirmed by our MOIRCS
   spectroscopy and \citet{hilton2010}.
   (b) Right panel: the colour-colour diagram of 
   $i'-z'_{\rm cont}$ vs. $R_c-i'$ for the \oii\ emitters and the other
   NB912 emitters which are identified based on the $Bz'K$ diagram (see
   the left panel). Note that in this plot, \ha\ emitters at $z=0.39$ 
   have already been excluded based on $R_c-i'$ and $B-R_c$ colours.
   Symbols are the same as in the left panel. Solid lines show the criteria
   used in \citet{ly2007}.
}
 \label{fig;bzk}
\end{figure*}

\begin{figure*}
 \begin{center}
 \includegraphics[width=0.85\linewidth]{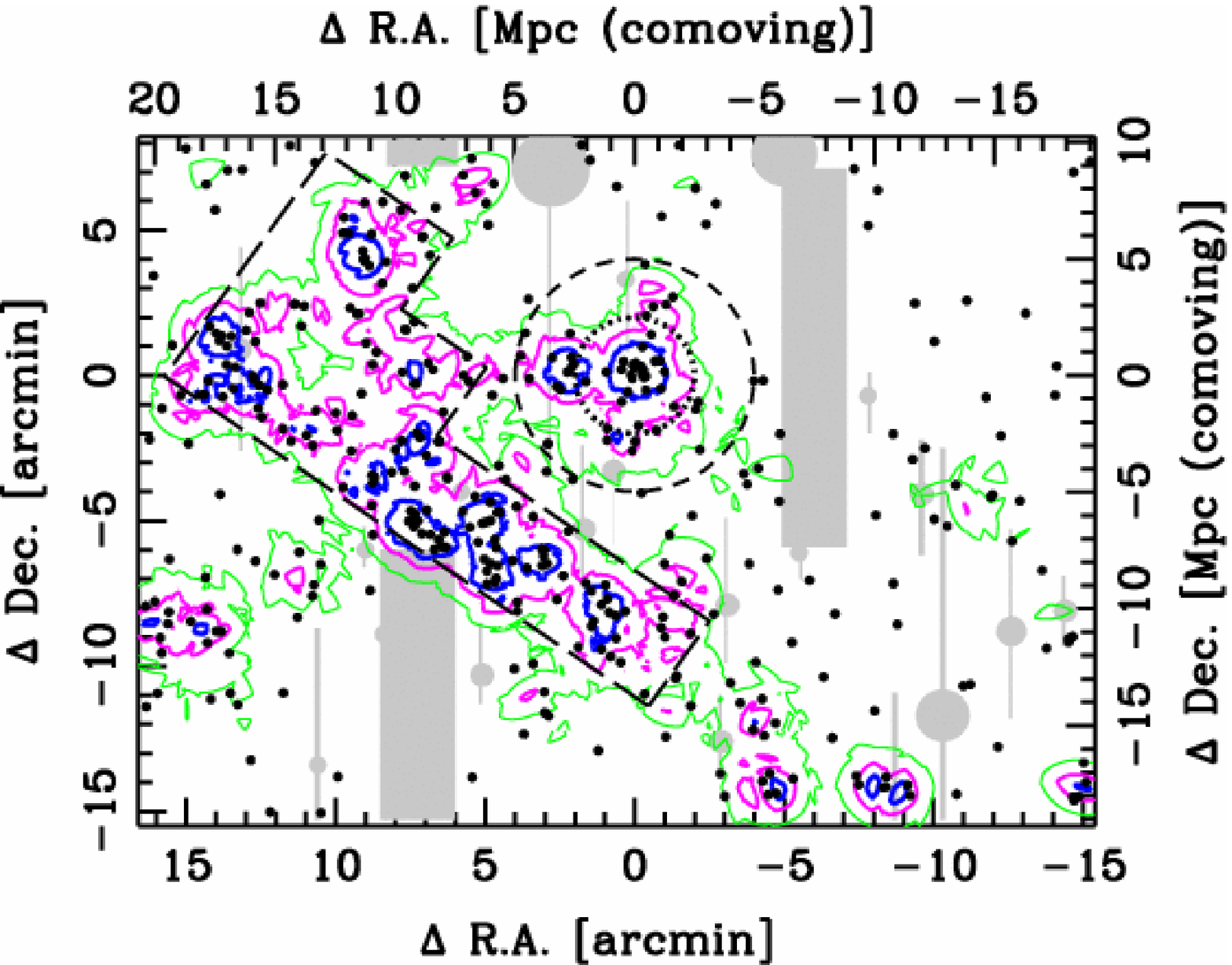}
 \end{center}
 \caption{The celestial distribution of our 380 \oii\ emitters at
   $z\sim1.46$ in and around the XCS2215 cluster. North is up, and east is
   to the left. The horizontal and vertical axes show relative
   coordinates from the cluster centre. Black dots show the \oii\
   emitters, and gray regions indicate the masked areas around bright stars
   which have bad quality image and were excluded from the analyses.
   Cluster core region is defined by a dotted-line circle with a
   radius of 2\arcmin, while outskirts region is defined as a ring
   with a width of 2\arcmin\ between the dotted and broken-line
   circles. Filament region shown by the long-dashed lines is defined
   to cover the prominent structure of the \oii\ emitters. 
   The rest of the area is defined as the field. 
   Blue, magenta and green contours show the local density of log\sigfifth[Mpc$^{-2}$]
   $=$1.07, 0.72, and 0.39, respectively (see \S\ref{sec;environment}).  
}
 \label{fig;map_oii}
\end{figure*}

We make a photometric catalogue in the same manner as
in \citet{hayashi2010}, but for the wide-field 32$\times$23 arcmin$^2$
data instead of the central 6$\times$6 arcmin$^2$ region of the
cluster. In this section, we briefly describe our updated photometric
catalogue.

Source detections are performed on the NB912 image using {\sc SExtractor}
\citep[ver. 2.5.0:][]{bertin1996}, and photometry on the other
images, except for the $J$-band, are conducted by the double-image mode of 
{\sc SExtractor}. Because the FoV of the $J$-band image is limited to the central
region and the pixel scale of MOIRCS image is smaller than that of
Suprime-Cam, the $J$ image is not matched to the NB912 image
geometrically. Therefore, we independently conduct source
detections and photometry on the $J$ image with {\sc SExtractor}, and 
cross-match the detected objects to those in the NB912-detected catalogue
if the coordinates of the $J$-detected objects are in agreement with
those of the NB912-detected objects within a 1\arcsec-diameter circle. 
Colour indices are derived from the 2\arcsec-diameter aperture magnitudes,
and {\sc mag\_auto} magnitudes are used as total magnitudes. Magnitude
errors are estimated from 1$\sigma$ sky noise taking account of the
difference in depth at each object position due to slightly different
exposure times and sensitivities.
Magnitudes are corrected for the Galactic absorption by the following
magnitudes; A($B$)=0.10, A($R_c$)=0.06, A($i'$)=0.05, A($z'$)=0.04,
A(NB912)=0.04, A($J$)=0.02, and A($K$)=0.01, which are derived from
the extinction law of \citet{cardelli1989} on the assumption of $R_V=3.1$ and
$E(B-V)=0.025$ estimated from \citet{schlegel1998}.
We check the zero-points of magnitudes in all the bands by comparing
stellar colours with those of stellar spectrophotometric atlas of
\citet{gunn1983}. The zero-point magnitudes are corrected so that
stellar colours are in good agreement with those of the stellar
atlas. Note that the correction is smaller than 0.15 magnitude at most.   

As a result, the catalogue contains 31,144 objects brighter than 25.20
mag. in NB912 (5$\sigma$ limiting magnitude) over the whole 32$\times$23
arcmin$^2$ region, except for the masked regions.
Among them, 27,430 galaxies are distinguished from 3,714 stars based
on $B-z'$ and $z'-K$ colours. This technique was devised by \citet{daddi2004},
and stars are actually well separated from galaxies on this colour-colour
diagram ($B-z'$ vs.\ $z'-K$) \citep{daddi2004,kong2006}.

\vspace{11mm}

\section{Selection of \oii\ emitters}
\label{sec;select-oii}

We already surveyed \oii\ emitters in the central region of the
XCS2215 cluster at $z=1.46$ with a combination of $NB912$ and $z'$-bands,
and identified 44 \oii\ emitters \citep{hayashi2010}. 
In this paper, we have expanded the survey area to the outskirts of
the cluster applying the same technique used in \citet{hayashi2010}.
We thus only describe a summary of our selection method below. 

First, we select galaxies with a emission line at $\sim$9139\AA\ by
applying the following criteria; 
\begin{enumerate}
\item $z'-NB912 \gid -2.5\log(1-\sqrt[]{\mathstrut f^2_{3\sigma,z'}+f^2_{3\sigma,NB912}}/f_{z'})$,
\item $z'-NB912 \gid -0.2$, 
\end{enumerate}
where $f_{3\sigma}$ is the 3$\sigma$ sky noise flux in each band and
$f_{z'}$ is the $z'$-band flux (Fig.~ \ref{fig;zznb}). The first
criterion is intended to select galaxies with an excess in $z'-NB912$ colour
redder than the 3$\sigma$ photometric error. This corresponds to a line flux
larger than 1.4$\times$10$^{-17}$ erg s$^{-1}$ cm$^{-2}$. 
If the excess is due to a \oii\ emission line at $z=1.46$, the limiting flux
corresponds to a dust-free SFR of 2.6 \Msun\ yr$^{-1}$ according to the
\oii--SFR calibration in \citet{kennicutt1998}.   
The second criterion corresponds to the observed equivalent width larger than
35\AA, which can exclude the possible contamination of galaxies
due mainly to photometric errors.
As discussed in \citet{hayashi2010}, the colour term in $z'-NB912$ is
negligible.  
As a result, we select 721 NB912 emitters from 27,430 galaxies over a
$\sim$700 arcmin$^2$ area (Fig.~\ref{fig;zznb}). Among them, 482
emitters are detected in $K$ at more than 2$\sigma$ level. In what
follows, we deal with the $K$-detected emitters, because we identify
the \oii\ emitters based on their $B-z'$ and $z'-K$ colours. This
means that our emitter sample is both flux- and mass-limited.

It is possible that the detected emission lines are any of the
following major strong lines;  
\ha\ at $z$=0.39, \oiii\ at $z$=0.82--0.84, \hb\ at $z$=0.88, and
\oii\ at $z$=1.46. 
Note that we find no candidates for Ly$\alpha$ emitters at $z$=6.51
in our sample, because our 482 NB912 emitters are all detected in
$i'$-band.
In order to discriminate \oii\ emitters from other lines at different
redshifts, we apply the colour selection criteria to the NB912 emitters;
\begin{equation} 
(z'_{\rm cont}-K)>(B-z'_{\rm cont})-0.46\ \cup\ (z'_{\rm cont}-K)>2.2,
\label{eq;member}
\end{equation}
where $z'_{\rm cont}$ corresponds to a continuum flux calculated from
equation (\ref{eq;fcont}) in \S\ref{sec;sfr_ssfr}.
The original idea comes from \citet{daddi2004} who devised the BzK
colour selection technique which can efficiently take out galaxies
located between $1.4<z<2.5$. We modified the original selection
boundaries in \citet{hayashi2010} so that we can sample galaxies at
$z=1.46$ more completely.
Note that the selection criteria (1) used in
this paper are yet slightly different from those adopted in
\citet{hayashi2010}. 
Firstly, we use $z'_{\rm cont}$ instead of
$z'$, which enables us to remove a contribution of emission line
to the continuum colour to be used in the colour selection.
The other is that we use $K$ magnitudes
taken with WFCAM/UKIRT instead of $K_s$ magnitude with 
MOIRCS/Subaru.
We estimate $K_s$(MOIRCS)$-K$(WFCAM) colours using spectral templates of
\citet{coleman1980} which are redshifted to 1.46, and find that the
difference of the filters can result in $K_s$(MOIRCS)$-K$(WFCAM)=0.02--0.05.
This works in the sense that the emitters close to the border tend to meet
the criteria more easily.

\begin{figure}
 \begin{center}
 \includegraphics[width=\linewidth]{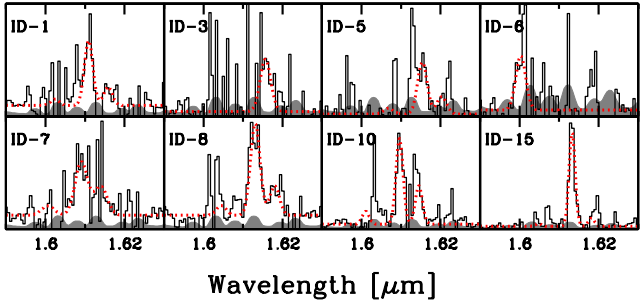}
 \end{center}
 \caption{ Spectra of the \oii\ emitters which show \ha\ and \nii\ emission
   lines, except for ID-3 and ID-6 where only \ha\ lines are detected.
   They cover a wavelength range of 1.59--1.63\micron\ in
   the observed frame. The vertical axis is flux density in arbitrary unit.     
   The solid lines show the spectra, and the gray regions show $1\sigma$
   errors.  Gaussian profiles are fitted to the lines and are shown by
   the red dotted lines together with the continuum levels.
}
 \label{fig;spectra_hanii}
\end{figure}

With the above criteria, we select 376 \oii\ emitters in total (Fig.~\ref{fig;bzk}(a)).
In \citet{hayashi2010}, we identified 44 \oii\ emitters in the central region
of the cluster. We check whether these previous sample of \oii\ emitters
are reproduced in the current sample in this paper.
Among the 44 \oii\ emitters, however, 11 objects are not identified as
\oii\ emitters in the current sample.
This is because of the update of photometric catalogue in particular in the $K$-band.
The MOIRCS $K_s$ image used in \citet{hayashi2010}
is $\sim0.3$ mag.\ deeper than the WFCAM $K$ image, and 
we miss some \oii\ emitters with faint $K$ magnitudes.
In fact, seven objects out of 11 are not detected in the WFCAM $K$ imaging.
However, for an uniformity of the data across the entire field, we use
the WFCAM data in this paper.
\citet{hilton2010} have spectroscopically confirmed 44 member galaxies
in the XCS2215 cluster.  We have also confirmed membership for
16 \oii\ emitters with spectroscopy (see \S~\ref{sec;spec-confirm}).
Among them, four NB912 emitters do not meet our \oii\ selection
criteria but they turn out to be real members at $z\sim1.46$.
We thus add them to our \oii\ emitter sample.
Consequently, our final \oii\ emitter sample consists of 380 galaxies. 

\citet{ly2007} have classified emission line galaxies using only the optical
colours. We test this selection method based on $R_c-i'$ vs.~$B-R_c$
and $i'-z'$ vs.~$R_c-i'$ diagrams used in \citet{ly2007} and check
whether it is effective in picking out \oii\ emitters at $z\sim1.46$.
If this classification works, we could select \oii\ emitters among those
faint in $K$ magnitude, i.e. less massive star-forming galaxies.
Fig.~\ref{fig;bzk}(b) shows $i'-z'$
and $R_c-i'$ colours for \oii\ and the other NB912 emitters identified
by the $Bz'K$ selection method, where \ha\ emitters at $z=0.39$ have already
been excluded from the NB912 emitter sample based on the
$R_c-i'$ vs.~$B-R_c$ colours \citep{ly2007}. 
In Fig.~\ref{fig;bzk}(b), it seems that \oii\ emitters and
the spectroscopically confirmed members are relatively well distinguished and
confined mostly in the bottom-right side of the diagram \citep{ly2007}.
However, we also note that there are many contaminations from other lines at
different redshifts.
This suggests that the NIR data is essential to photometrically select galaxies
at $z\sim$1--2, and we decide not to use the optical colour selection.

Also, we calculate photometric redshifts using the {\sc eazy} code
\citep{brammer2008} with six SED templates of EAZY\_v1.0 and based on
the photometry in $B, R_c, i', z'$ and $K$ bands for the $K$-detected
subsample of spectroscopic members.
However, 20\% of the spectroscopic members have completely wrong photometric
redshifts at $z_{\rm phot}<1$. Therefore we do not use the photometric redshift
technique either.

Therefore, in this paper, we will rely on the $Bz'K$ colour selection
in identifying the \oii\ emitters at $z\sim1.46$ among the NB912 emitters.

Fig.~\ref{fig;map_oii} shows the distribution of \oii\ emitters at
$z\sim1.46$, where gray regions are masked due to bad quality of the
image. We confirm that many \oii\ emitters exist in the central 
region, as is already reported in \citet{hayashi2010}. Moreover,
due to the extension of our survey to the outskirts of the XCS2215
cluster, we also find that there is a prominent filamentary
large-scale structure of \oii\ emitters from the east to the south of
the cluster (Fig.~\ref{fig;map_oii}). 
This filament is surely one of the largest structures of star-forming galaxies
at $z=1.46$. Recent studies suggest that at lower redshifts
medium-density regions embedded in large-scale structures surrounding
clusters are crucial sites to understand galaxy evolution 
\citep{tanaka2005,koyama2008}. The discovery of the large-scale
structure at $z=1.46$ provides us with a unique opportunity to investigate
environmental dependence of galaxy properties in particular those in such
interesting medium-density environments at this high redshift.
In \S\ref{sec;phot-properties}, we examine the properties of galaxies in the
filamentary structure, and compare them with those in different environments.  

\begin{table*}
 \caption{A catalogue of the spectroscopically confirmed \oii\ emitters.
  }
\begin{center}
\begin{tabular}{cccccccccccccc}
\hline\hline
ID & Slitmask & R.A. & Dec. & Redshift & \multicolumn{3}{c}{Magnitude and Colour$^\dagger$} && \multicolumn{5}{c}{Emission lines$^\ddagger$} \\
\cline{6-8}\cline{10-14}\\[-1.5mm]
 & & & & & $K_s$ & $B-z'$ & $z'-K_s$ && \oii & \hb & \oiii & \ha & \nii \\
\hline
1 & MOS1 & 22:15:57.22 & -17:37:53.25  & 1.455 &  21.20 & 1.91 & 2.04 &   & --- & $\times$ & $\times$ & $\circ$ & $\circ$ \\ 
2 & MOS1 & 22:15:58.84 & -17:38:10.07  & 1.455 &  20.91 & 1.37 & 1.67 &   & --- & $\times$ & $\circ$ & $\times$ & $\times$ \\ 
3 & MOS1 & 22:15:52.89 & -17:39:08.10  & 1.462 &  20.97 & 1.09 & 1.24 &   & --- & $\times$ & $\times$ & $\circ$ & $\times$ \\ 
4 & MOS1 & 22:16:09.00 & -17:38:32.66  & 1.460 &  22.01 & 1.03 & 1.03 &   & $\circ$ & $\circ$ & $\circ$ & --- & --- \\ 
5 & MOS1 & 22:16:02.54 & -17:37:56.45  & 1.461 &  22.51 & 1.47 & 1.60 &   & $\times$ & $\times$ & $\times$ & $\circ$ & $\circ$ \\ 
6 & MOS1 & 22:15:56.86 & -17:35:58.52  & 1.440 &  22.86 & 0.89 & 0.94 &   & $\circ$ & $\times$ & $\circ$ & $\circ$ & $\times$ \\ 
7 & MOS2 & 22:15:57.69 & -17:37:45.77  & 1.452 &  20.22 & 2.50 & 2.72 &   & --- & $\circ$ & $\times$ & $\circ$ & $\circ$ \\ 
8 & MOS2 & 22:16:02.40 & -17:39:53.35  & 1.460 &  20.98 & 1.31 & 1.54 &   & $\circ$ & $\circ$ & $\circ$ & $\circ$ & $\circ$ \\ 
9 & MOS2 & 22:15:57.39 & -17:37:04.11  & 1.461 &  21.47 & 1.42 & 1.20 &   & $\circ$ & $\times$ & $\circ$ & $\times$ & $\times$ \\ 
10 & MOS2 & 22:15:57.18 & -17:38:07.88  & 1.454 &  21.68 & 1.15 & 1.28 &   & --- & $\circ$ & $\circ$ & $\circ$ & $\circ$ \\ 
11 & MOS2 & 22:16:07.43 & -17:37:25.22  & 1.461 &  21.99 & 1.22 & 1.08 &   & $\circ$ & $\times$ & $\circ$ & --- & --- \\ 
12 & MOS2 & 22:16:07.63 & -17:36:35.68  & 1.465 &  22.51 & 0.71 & 0.58 &   & $\circ$ & $\circ$ & $\circ$ & --- & --- \\ 
13 & MOS2 & 22:15:56.90 & -17:38:33.88  & 1.448 &  22.64 & 0.63 & 1.51 &   & $\circ$ & $\circ$ & $\times$ & $\times$ & $\times$ \\ 
14 & MOS2 & 22:15:49.50 & -17:38:58.35  & 1.465 &  22.79 & 0.67 & 0.79 &   & $\circ$ & $\times$ & $\circ$ & --- & --- \\ 
15 & MOS2 & 22:15:59.46 & -17:38:37.61  & 1.460 &  22.80 & 0.71 & 0.26 &   & --- & $\circ$ & $\circ$ & $\circ$ & $\circ$ \\ 
16 & MOS2 & 22:16:08.05 & -17:37:55.95  & 1.461 &  23.19 & 0.78 & 0.78 &   & $\circ$ & $\times$ & $\circ$ & --- & --- \\ 
\hline\hline
\multicolumn{14}{l}{ $\dagger$ Magnitudes and colours are taken from the catalogue of \citet{hayashi2010}. }\\
\multicolumn{14}{l}{ $\ddagger$ $\circ$: detected, $\times$: not detected, ---: no data }
\label{tbl;spec_sample}
\end{tabular}
\end{center}
\end{table*}

\begin{figure}
 \begin{center}
 \includegraphics[width=\linewidth]{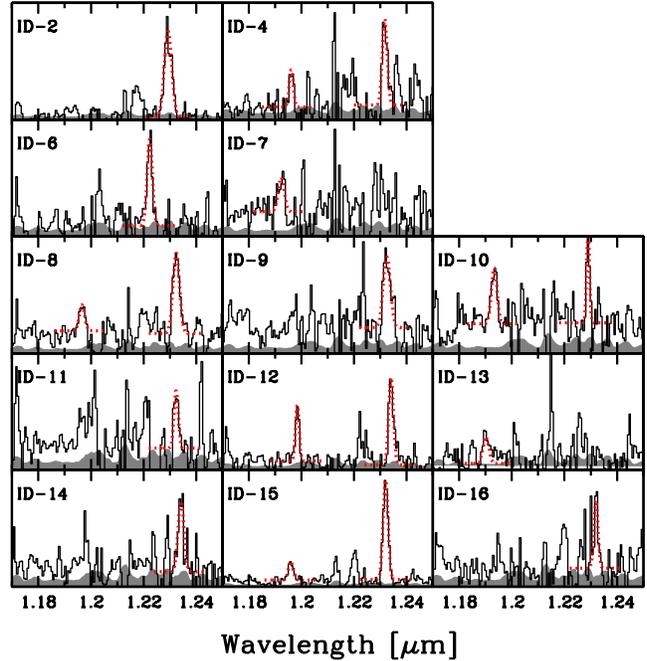}
 \end{center}
 \caption{ Same as Fig.~\ref{fig;spectra_hanii}, but for \hb\ and
 \oiii\ emission lines.  The spectra cover a wavelength range of
 1.17--1.25\micron\ in the observed frame. Gaussian profiles with 
 continuum levels are shown wherever the \hb\ and/or
 \oiii($\lambda$5007) are detected. 
}
 \label{fig;spectra_oiiihb}
\end{figure}

\section{Spectroscopic confirmation of the \oii\ emitters}
\label{sec;spec-confirm}

In \citet{hayashi2010} and the previous section, we photometrically
identify 380 \oii\ emitters at $z=1.46$ in the XCS2215 cluster
and its outskirts. However, the confirmation of \oii\ emitters by
spectroscopy is essential to verify that our method is valid for the
selection of \oii\ emitters at $z=1.46$. 

By our MOIRCS spectroscopy, we have detected some emission lines, such
as \ha, \hb, \oiii, \nii, for 16 out of the 34 targeted \oii\ emitters
in the cluster region.
In the case of narrow-band emitters, it is relatively easy to identify
the detected emission line even if only a single line is detected on the
spectrum. Because we already detect a emission line at $\sim9139$\AA\ by
narrow-band imaging, the possible redshifts of the emitters are very
limited. For \oii\ emitters at $z=1.46$, \ha\ and \nii\ lines should
be seen at $\lambda\sim$1.61\micron, while \hb\ and \oiii\ lines should be seen
at $\lambda\sim$ 1.19\micron\ and 1.23\micron, respectively.

We first perform line detections in the two-dimensional spectra by
visual inspection, taking care not to be confused by residual OH airglow
lines or any accidental noises. Among the
confirmed lines, we then regard a line with a flux larger than 2$\sigma$
of the sky noise as a real signal.
Figs.~\ref{fig;spectra_hanii} and \ref{fig;spectra_oiiihb} show the
one-dimensional spectra of the lines, and a summary of the detected
lines for the 16 confirmed \oii\ emitters is shown in Table~\ref{tbl;spec_sample}.    

The spectroscopic redshifts and the fluxes of the emission lines are
measured
by fitting Gaussian profiles, where the free parameters are amplitude,
line width, and redshift. Before fitting, we estimate a constant
continuum level using the regions close to the line and without a
contribution of strong OH lines, and subtract it from each spectrum.    
The $1\sigma$ error of the spectrum is estimated from the covariance of
$\chi^2$ fitting based on the sky noise. We confirm that the error
is comparable to the sum of the sky noise within 2$\times$FWHM
around each line. If a \nii\ emission line is seen in a spectrum,
\ha\ and a doublet of \nii\ ($\lambda\lambda6548,6584$) are simultaneously
fit to the spectrum, assuming the same line width and redshift, and a 
\nii($\lambda6584$)/\nii($\lambda6548$) ratio of 3. 
In the case where several lines are detected for an emitter, we determine
its redshift by taking an average of the redshifts measured by individual lines.

\begin{figure}
 \begin{center}
 \includegraphics[width=\linewidth]{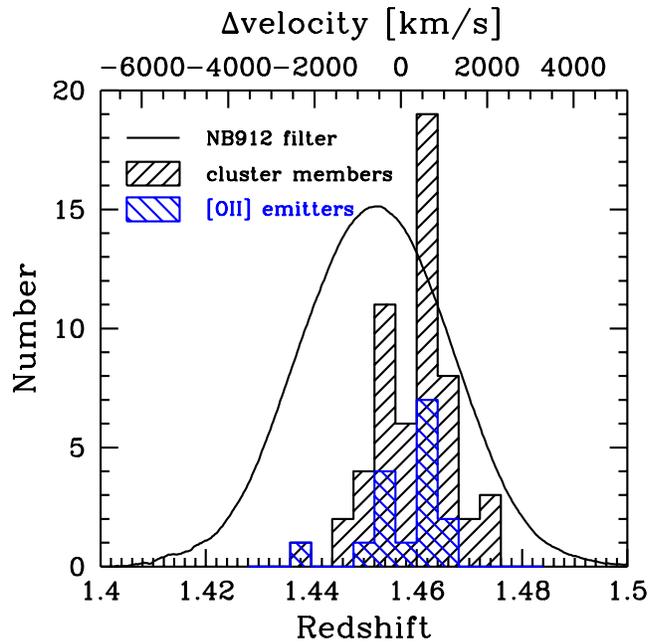}
 \end{center}
 \caption{Redshift distributions of the cluster member galaxies.
 The blue histogram shows the 16 \oii\ emitters which are spectroscopically
 confirmed by our MOIRCS observations, while the black histogram shows all the
 56 members of which 40 redshifts are taken from \citet{hilton2010}.
 Among the 44 members in \citet{hilton2010}, four objects overlap with
 our \oii\ emitters that are spectroscopically confirmed with MOIRCS. 
 The solid curve shows the response function of the NB912 filter in an
 arbitrary unit.
 }
 \label{fig;redshift}
\end{figure}

\begin{figure}
 \begin{center}
 \includegraphics[width=\linewidth]{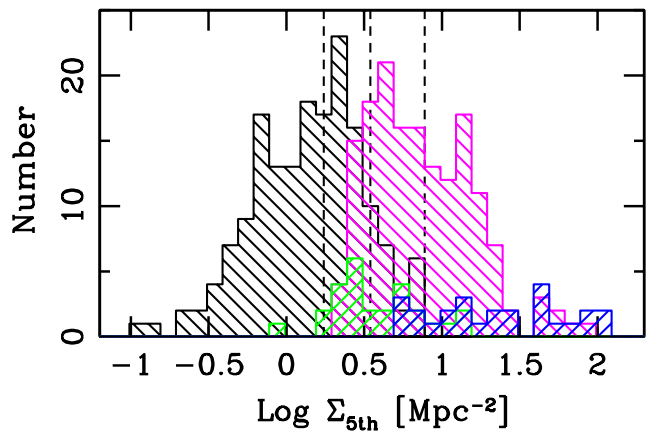}
 \end{center}
 \caption{
   The histograms show the numbers of \oii\ emitters as a function
   of the local density, \sigfifth. Blue, green, magenta, and black
   histograms correspond to the core, outskirts, filament, and
   the field regions, respectively (see also Fig.~\ref{fig;map_oii}).  
   The three vertical dashed lines show the boundaries between the
   four environments defined based on the local density. 
}
 \label{fig;region}
\end{figure}

As a result, we confirm that 16 \oii\ emitters are certainly
cluster member galaxies at $z\sim1.46$, which is nearly 50 per cent
(16/34) of our \oii\ emitter sample.
For other three targets, we do see an emission line at $\sim9100$\AA,
but it is the only detected line and we are not sure whether it is an \oii\ line.
The spectra of two of them do not cover the wavelength where \ha\ line
should be seen, unfortunately.
It is also possible that the detection limit of \ha\ at
$\sim$1.61\micron\ is shallower than that of \oii\ at $\sim0.91$\micron.
For the remaining 15 targets, no emission line is
detected even if a spectrum covers the wavelength range down to 0.9\micron.
In MOS1 (MOS2) mask, emission lines are detected for targets with \oii\
fluxes larger than $0.41 (0.23) \times10^{-16}$ erg s$^{-1}$
cm$^{-1}$ (see Table \ref{tbl;line-ratio}).
We find that 12 out of the 15 targets without any line detection have
estimated \oii\ fluxes similar to or smaller than the limiting flux.
For the two targets with estimated \oii\ fluxes large enough
to be detected, their spectra do not cover the wavelength range down to
0.9\micron\ and we cannot confirm their \oii\ emission lines.
They do cover, however, the wavelength range up to 1.7\micron\ where
\ha\ lines are expected to show up. But we do not detect them, either.
For the remaining one target, although we obtain its spectrum covering
0.9--1.35\micron, no line is detected.

Therefore, except for the three targets with strong enough estimated
\oii\ fluxes, non-detection of any emission lines for the \oii\ emitter
candidates are probably because of intrinsically too weak \oii\ emission lines,
as well as other lines which are also weaker than the detection limit.
Non-photometric weather condition during our spectroscopy
may have resulted in reducing the success rate of line detections.

Fig.~\ref{fig;redshift} shows the redshift distribution of the confirmed
\oii\ emitters and that of all the cluster members.
We cross-identify our 16 \oii\ emitters to the 44 member galaxies in
\citet{hilton2010}, and four galaxies turn out to be common objects
whose coordinates match together to an accuracy of 1$\arcsec$.
\citet{hilton2010} suggest that the redshift distribution may have
double peaks, and that such profile may indicate that this cluster
experienced a cluster-cluster merger event within the past few Gyrs.
A similar double-peaked profile is also seen for the \oii\ emitters
as well as all the member galaxies (Fig.~\ref{fig;redshift}). 
Although \citet{hayashi2010} found a high fraction of \oii\ emitters
in the core region of the XCS2215 cluster, we were not able to
reject the possibility that it is due to a projection effect of
the \oii\ emitters in the outskirts along the line of sight 
which are apparently superposed onto the cluster core.
However, similarity of the redshift distributions between the \oii\
emitters and all the member galaxies (Fig.~\ref{fig;redshift})
strongly suggests that the emitters are indeed located in the cluster
core in space.

\begin{figure*}
 \begin{center}
 \includegraphics[width=0.48\linewidth]{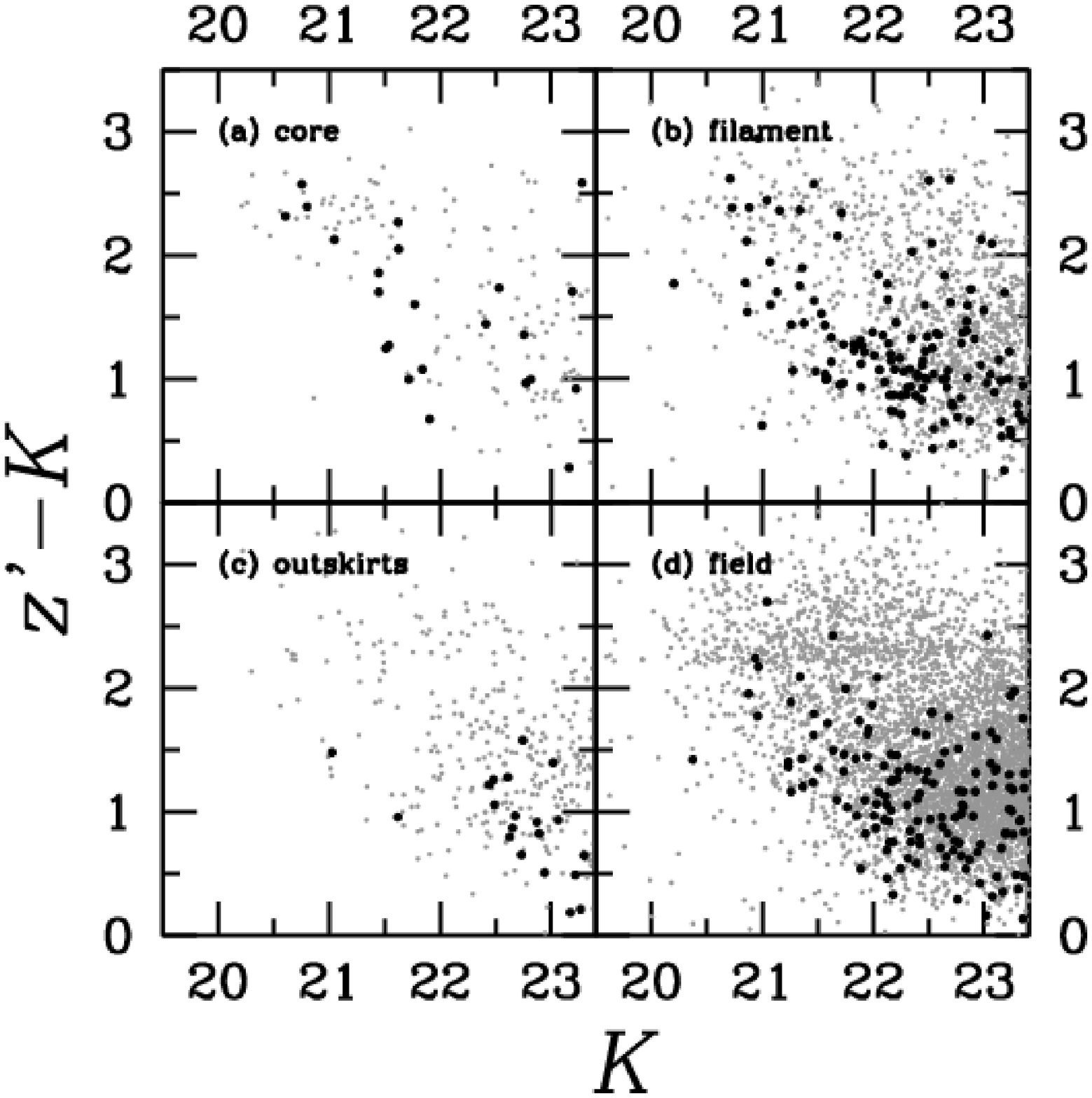}
 \includegraphics[width=0.48\linewidth]{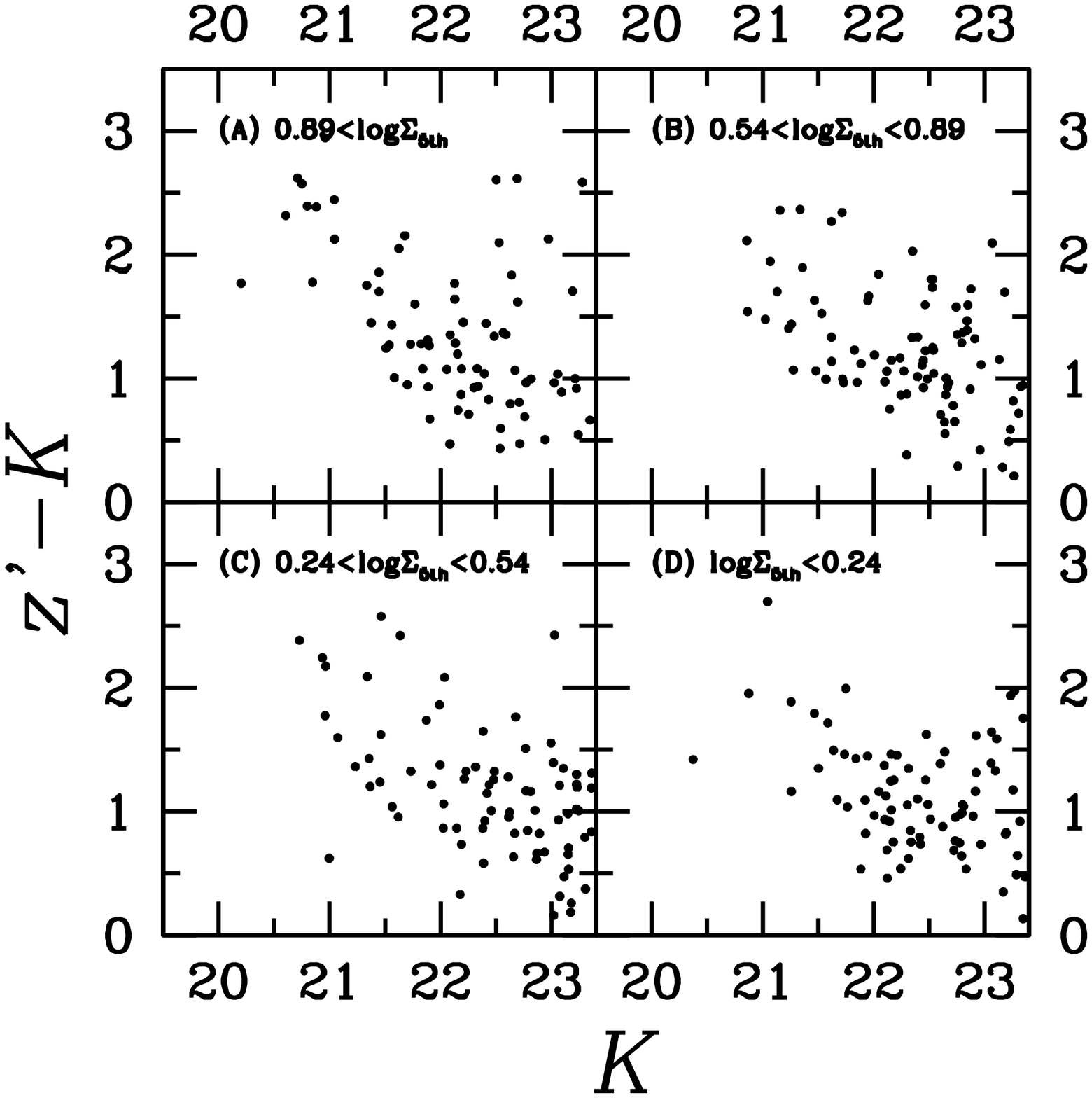}
 \end{center}
 \caption{ The $z'-K$ vs. $K$ colour--magnitude diagram of \oii\
   emitters in each environment.
   (Left panel) Black dots are the \oii\ emitters, while gray dots are
   the galaxies which meet colour criteria given by the equation (\ref{eq;member}). 
   (Right panel) Same as the left panel, but the environments are separated based on
   the local density of \sigfifth; 
   (A) high density region, 0.89$<$$\log$\sigfifth, 
   (B) medium-high density region, 0.54$<$$\log$\sigfifth$<$0.89, 
   (C) medium-low density region, 0.24$<$$\log$\sigfifth$<$0.54, and 
   (D) low density region, $\log$\sigfifth$<$0.24. 
   See text in \S\ref{sec;environment} for the details of the four
   environments. Only the \oii\ emitters are plotted in this panel,
   because the local density is calculated with the \oii\ emitter
   sample.  
}
 \label{fig;cmd}
\end{figure*}

\section{Environmental dependence} 
\label{sec;phot-properties}

\subsection{Definition of environments}
\label{sec;environment}

We define four different environments (core, outskirts, filament, and the field)
based on the spatial distribution of \oii\ emitters as shown in Fig.~\ref{fig;map_oii}.
The core is defined as the circled region with a radius of 2\arcmin\ from the cluster centre,
while the outskirts is defined as the ring with an inner radius of 2\arcmin\ and an
outer radius of 4\arcmin. The filament is defined as the enclosed area by the long-dashed
lines which neatly cover the notable filamentary structure characterized
by relatively high density of the \oii\ emitters.
All the rest is defined as the field region.

As another definition of environment, we will also use local density \sigfifth, which is
calculated using the area where the fifth nearest \oii\ emitters are
included. In the local density measurements,
it would be desirable to use a whole sample of $z=1.46$ galaxies including
not only blue star-forming galaxies but also red quiescent galaxies.
However, the current photometric redshifts
(without $J$-band in particular) are not accurate enough to define such local
density of the whole population.
Therefore, we count only the \oii\ emitters to define local density
as they are most likely located at $z=1.46$ with little contamination from foreground or
background galaxies.
It should be noted that \citet{hayashi2010} found that the fraction
of \oii\ emitters in this cluster is almost constant irrespective of
the distance from the cluster centre. This suggests that even if only
\oii\ emitters are used to calculate the local density, we can expect to
trace the structure of the whole population to some extent.

Fig.~\ref{fig;region} shows the distribution of local density
\sigfifth\ of the \oii\ emitters in each environment defined above
based on Fig.~\ref{fig;map_oii}. If the four environments are
rearranged in the order of decreasing local density, the core is the
highest density region, filament is the second, the outskirts
is the third, and the field is the lowest density regions. Note that
the large-scale filamentary structure (i.e.\ filament) is actually
denser than the cluster outer region (i.e.\ outskirts). 

When we discuss environmental dependence of galaxy properties below
based on the local density, we divide the galaxies into four classes
according to the local densities, each of which contains an equal number
of \oii\ emitters ($\sim95$). The boundaries between the classes are shown
by the two vertical dotted lines in Fig.~\ref{fig;region}.

\subsection{colour--magnitude diagram}
\label{sec;cmd}

Fig.~\ref{fig;cmd} shows the colour--magnitude diagrams in each
environment. In the left panel the four environments are defined
based on the spatial distribution of the \oii\ emitters,
while in the right panel the four environments are divided based on
the local density (see \S\ref{sec;environment}).  
The distribution of the \oii\ emitters on the colour--magnitude
diagrams indicates that the \oii\ emitters tend to be blue galaxies
in general as expected. However, there are some red \oii\ emitters
with $z'-K>2.2$.  The fraction of the red \oii\ emitters seems higher
in high density regions, such as core, filament, and high-\sigfifth\
regions, compared to other environments. 

\begin{figure}
 \begin{center}
 \includegraphics[width=\linewidth]{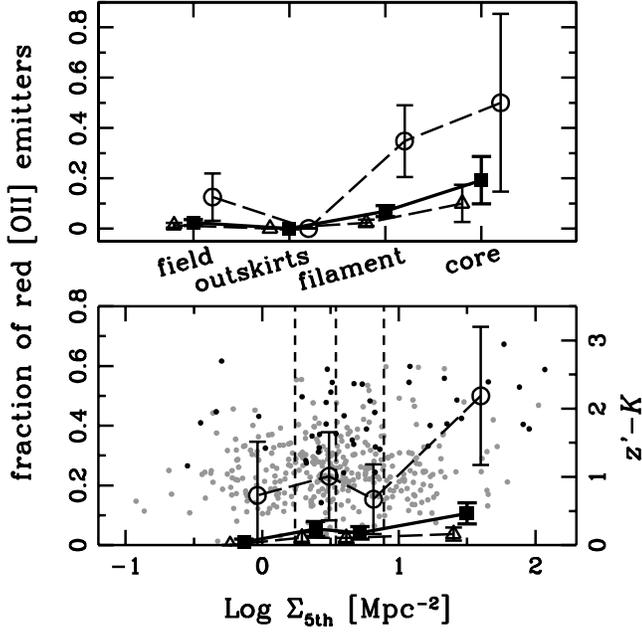}
 \end{center}
 \caption{Fraction of the red \oii\ emitters with $z'-K>2.2$ to all the 
   \oii\ emitters is shown by the filled squares as a function of
   environment. Open circles and open triangles, connected by the
   long-dashed lines, show the fractions separated into bright and faint
   sample at $K=21.5$, respectively.
   The data points are slightly offset manually to avoid their overlapping.
   The four environmental bins defined in \S\ref{sec;environment} are used 
   in the upper panel, while the local densities are used in the lower
   panel. In the lower panel, the black dots show the \oii\ emitters
   with $K<21.5$ and the gray dots show those with $K>21.5$. Four
   density bins are separated by the vertical dashed lines. 
   The right vertical axis shows $z'-K$ colours of the \oii\ emitters. 
}
 \label{fig;ref_oii_frac}
\end{figure}

Fig.~\ref{fig;ref_oii_frac} shows this trend quantitatively. 
The fraction of the red \oii\ emitters ($z'-K>2.2$) to all the \oii\
emitters is plotted as a function of environment. The galaxies redder
than $z'-K=2.2$ can be called as ``red-sequence'' galaxies for
the cluster redshift at $z=1.46$ \citep{hayashi2010}. The figure
clearly shows an excess of red \oii\ emitters towards high(-est)
density regions. \citet{koyama2010} reported that such red
star-forming galaxies in the RXJ1716.8+6708 cluster at $z=0.81$ are
preferentially seen in medium-density regions, such as groups,
filaments, and the outskirts of the cluster, rather than in
high-density cluster core. They claim that it indicates that the
cluster at $z=0.81$ has already quenched the star formation activity
in the high-density region, and the region of active star formation
has been shifted to the medium-density regions. \citet{tanaka2009}
also found that there are few red galaxies with \oii\ emission in the
RDCS~J1252.9-2927 cluster at $z=1.24$, and such galaxies tend to exist
in groups or in the field. These facts may suggest that the central
region of the cluster at $z=1.46$ is at a similar stage of galaxy
evolution to the outskirts of clusters at lower redshifts, where red
star-forming galaxies appear.   

It is clear from Fig.~\ref{fig;cmd}, that the colour and magnitude of the
\oii\ emitters correlate in the sense that brighter emitters in $K$ tend
to be redder in $z'-K$. Furthermore, in this cluster, the
colour--magnitude diagram shows a deficit of red sequence galaxies
with $K$ fainter than $\sim$21.5 \citep{hayashi2010}. Galaxy
properties are thus strongly magnitude dependent. We therefore
investigate the dependence of the fraction of red \oii\ emitters on
the $K$-band luminosity. It is found that the higher fraction of red
\oii\ emitters towards high density regions is dominated by massive
\oii\ emitters with $K<21.5$ (Fig.~\ref{fig;ref_oii_frac}).  For less
massive \oii\ emitters with $K>21.5$, the fraction of red \oii\
emitters is very small, and is not strongly dependent on the
environment. 
It is likely that massive \oii\ emitters in high density regions change
to red colours earlier than less massive ones and/or in lower density
environments. This is consistent with the down-sizing scenario that
more massive galaxies become quiescent galaxies earlier, and
its environmental dependence in the sense that galaxy evolution (and
down-sizing) proceeds earlier in higher density regions \citep[e.g.,][]{tanaka2005}.

\begin{figure}
 \begin{center}
 \includegraphics[width=\linewidth]{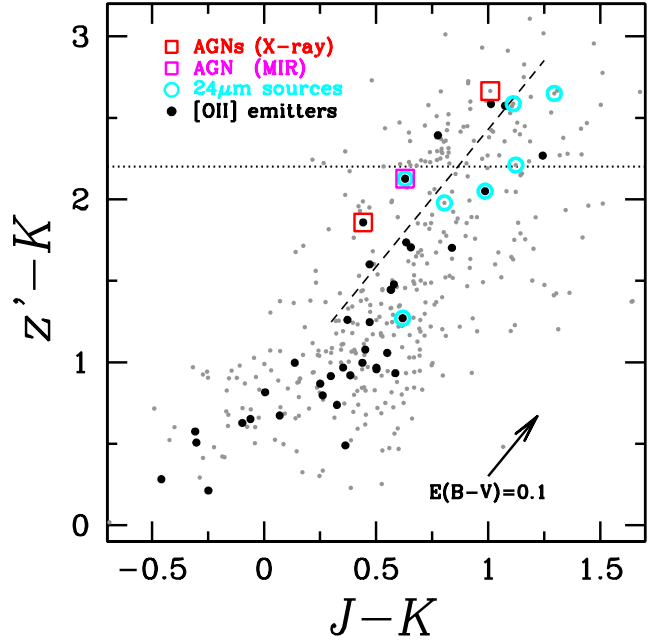}
 \end{center}
 \caption{The colour--colour diagram of $J-K$ vs.~$z'-K$ for the
   galaxies in the cluster centre where $J$-band photometry is
   available. Black dots show the \oii\ emitters, and gray dots are
   the galaxies which meet our $Bz'K$ criteria given by the equation
   (\ref{eq;member}). Red and magenta squares show AGNs identified by
   X-ray and mid-infrared colours, respectively \citep{hilton2010}. 
   Cyan circles show dusty starburst galaxies which are detected in
   24\micron\ by Spitzer/MIPS \citep{hilton2010}. The arrow shows a
   reddening vector corresponding to $E(B-V)=0.1$, which is estimated
   from the extinction curve of \citet{calzetti2000}. The dashed line
   drawn in parallel to the reddening vector approximately separates
   between passive galaxies (upper) and dusty galaxies (lower). 
   The dotted line shows the colour of $z'-K=2.2$.
}
 \label{fig;zjk_colour}
\end{figure}

Fig.~\ref{fig;zjk_colour} shows the $J-K$ vs.~$z'-K$ colour--colour
diagram for the $6'\times6'$ region in the cluster centre where
$J$-band photometry is available. We utilize this diagram to
investigate the nature of the red \oii\ emitters. 
This method is analogous to the one that is used to separate between
starburst galaxies and passive galaxies for Extremely Red Objects
(EROs) based on the strength of the Balmer/4000\AA\ break feature
\citep[e.g.,][]{pozzetti2000}. 
In the XCS2215 cluster, \citet{hilton2010} found eight 24\micron\
sources with Spitzer/MIPS and three AGNs identified by Chandra X-ray
data and Spitzer/IRAC mid-IR colours. Among them, we cross-identified
seven dusty starbursts and three AGNs in our catalogue of cluster
member candidates, and plot them together in the figure. The arrow
shows a reddening vector estimated from the extinction curve of
\citet{calzetti2000}. We can notice two sequences of galaxies on this
diagram separated by the dashed line which is drawn in parallel to the
reddening vector. One is for passively evolving galaxies with a redder
$z'-K$ colour for a given $J-K$, and the other is star-forming
galaxies which are distributed in the direction of the dust reddening
vector extended from the blue-cloud. The three confirmed AGNs are
distributed along the redder sequence for passive galaxies.
Interestingly, none of the red \oii\ emitters are detected in
24\micron, and most of them are located along the passive sequence
with confirmed AGNs. 
Therefore, it is likely that a high fraction of red \oii\ emitters in
massive galaxies in the cluster core is caused by enhanced AGN
activity in almost passively evolving galaxies.

\citet{yan2006} found that \oii\ emission lines from red
galaxies tend to be produced by AGN activities rather than star
forming activities in the local Universe. \citet{lemaux2010} also
found similar results at $z$=0.8--0.9. 
These support our results. The high fraction of red \oii\ emitters in
high density regions may suggest that AGN feedback is contributing to
quench star formation activities in galaxies. In fact, as we discuss
in \S\ref{sec;AGN}, the line ratio diagnosis indicates a moderate
level of AGN contribution in the \oii\ emitters located in the central
region of the XCS2215 cluster.   

\subsection{SFR, specific SFR, and stellar mass}
\label{sec;sfr_ssfr}

\begin{figure*}
 \begin{center}
 \includegraphics[width=0.48\linewidth]{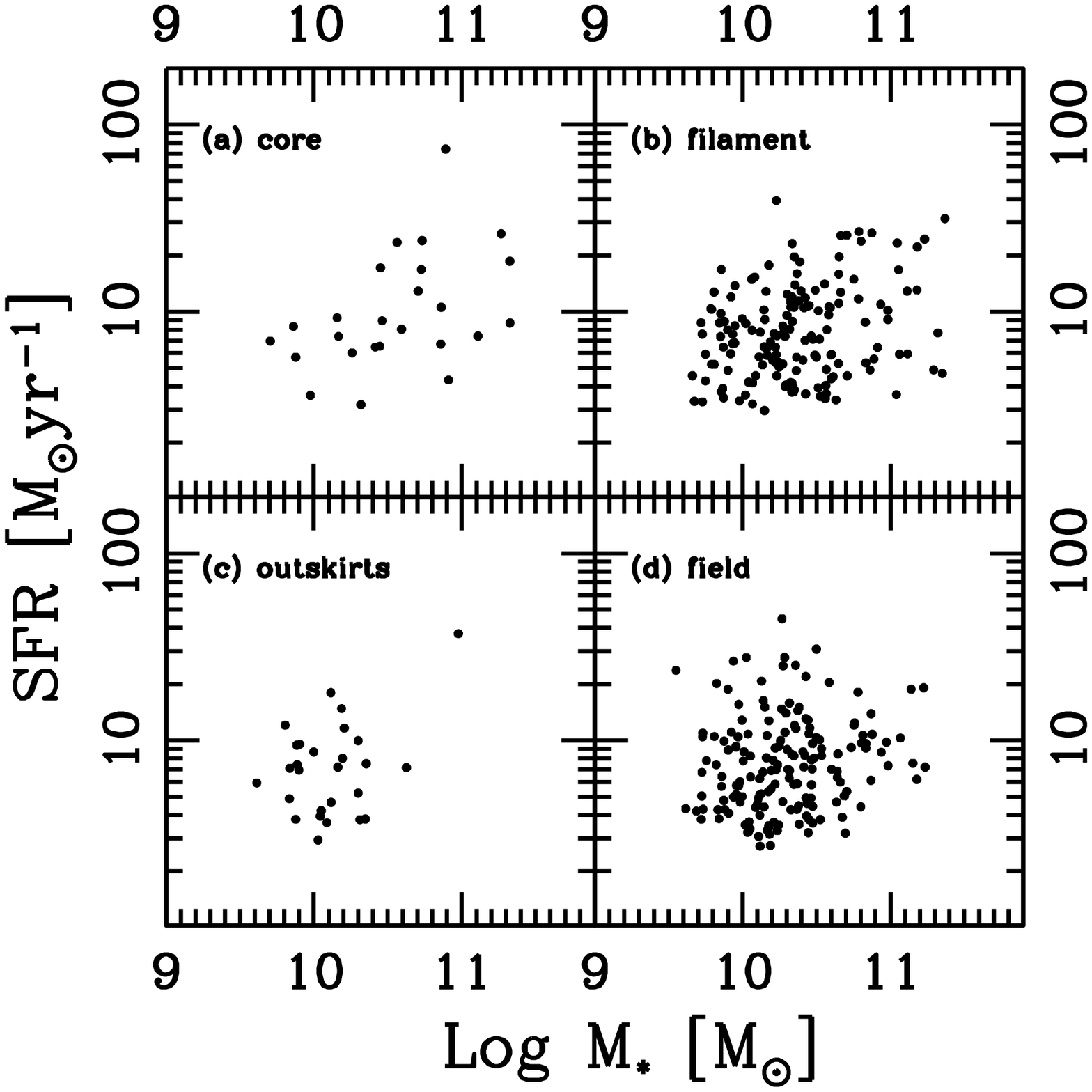}
 \includegraphics[width=0.48\linewidth]{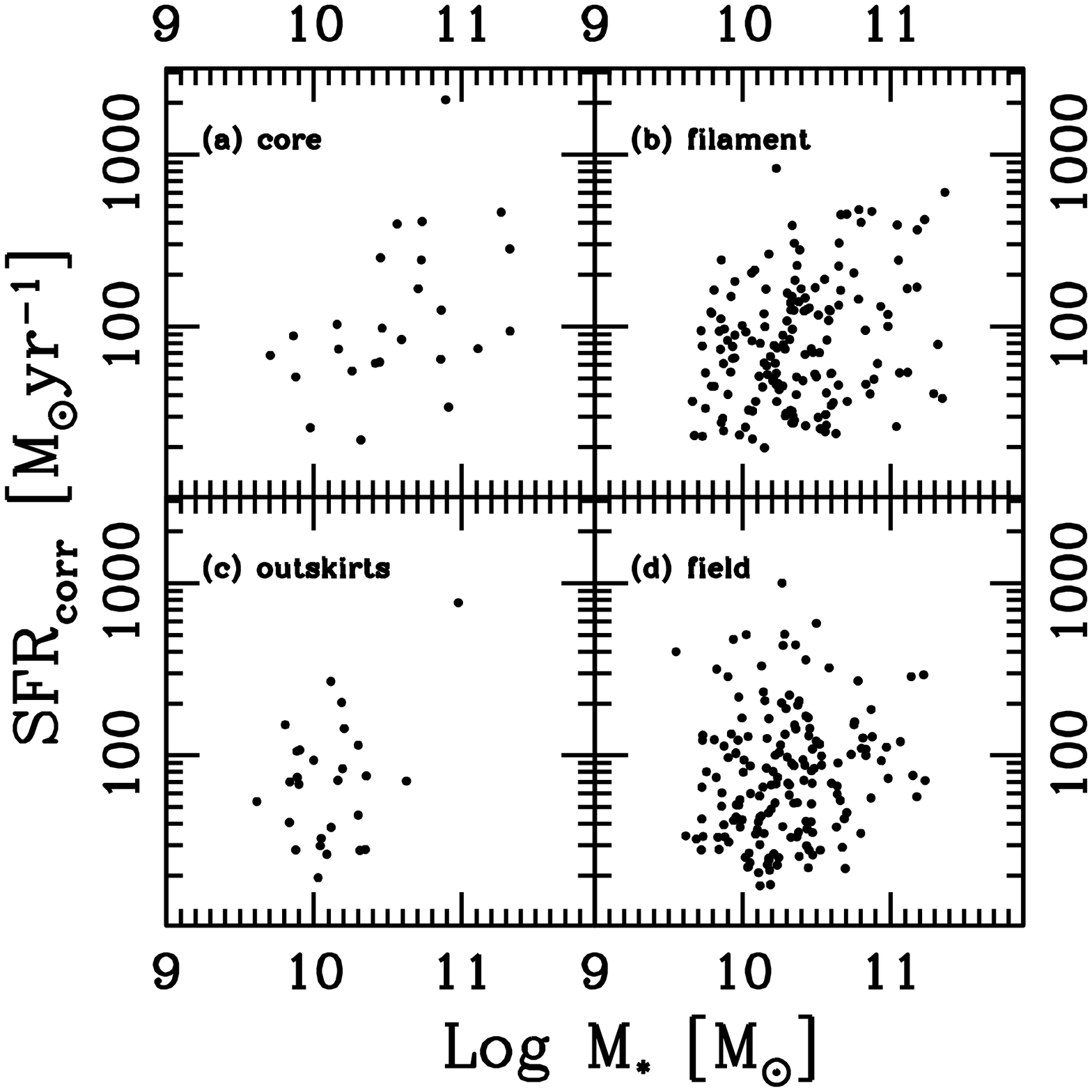}
 \end{center}
 \caption{SFRs of the \oii\ emitters as a function of stellar mass in four
   environments; (a) core, (b) filament, (c) outskirts, and (d) field. 
   The left panel shows the SFRs converted from the observed \oii\
   luminosities, while the right panel shows the dust-corrected
   SFRs. See text for the details of the dust correction.
}
 \label{fig;mass-sfr}
\end{figure*}

\begin{figure*}
 \begin{center}
 \includegraphics[width=0.48\linewidth]{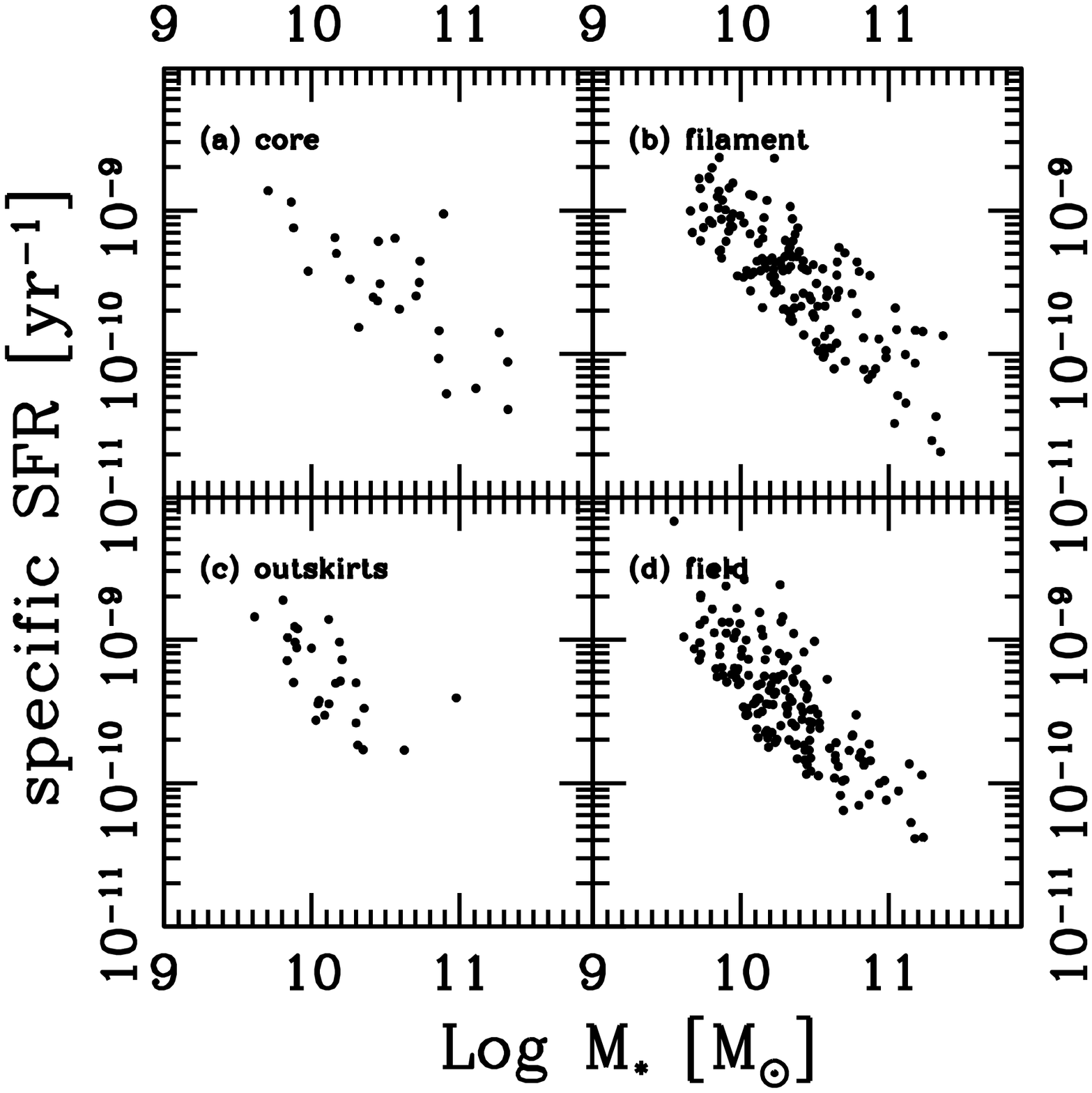}
 \includegraphics[width=0.48\linewidth]{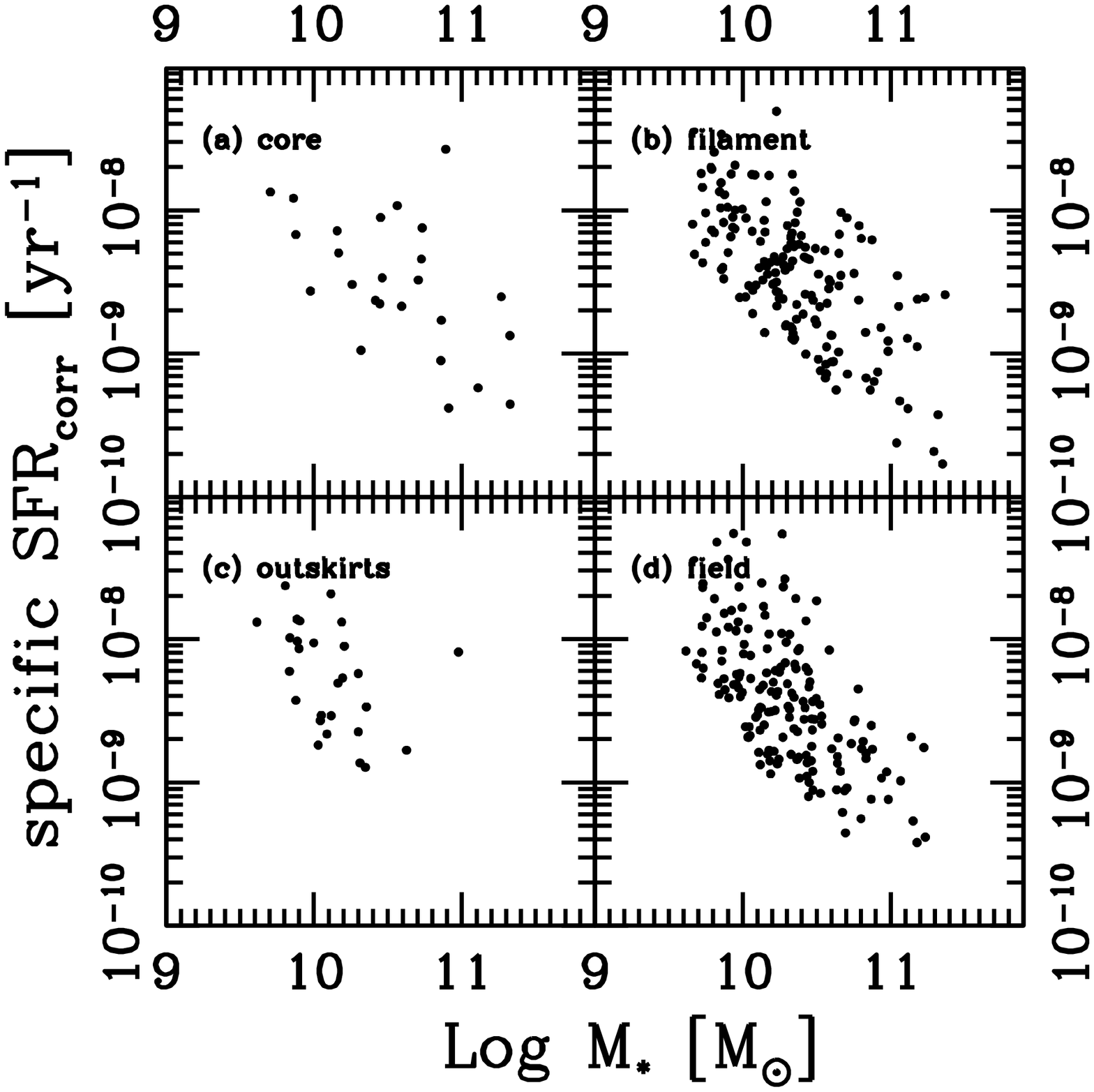}
 \end{center}
 \caption{Same as Fig.~\ref{fig;mass-sfr}, but for the specific SFRs.
}
 \label{fig;mass-ssfr}
\end{figure*}

\begin{figure}
 \begin{center}
 \includegraphics[width=\linewidth]{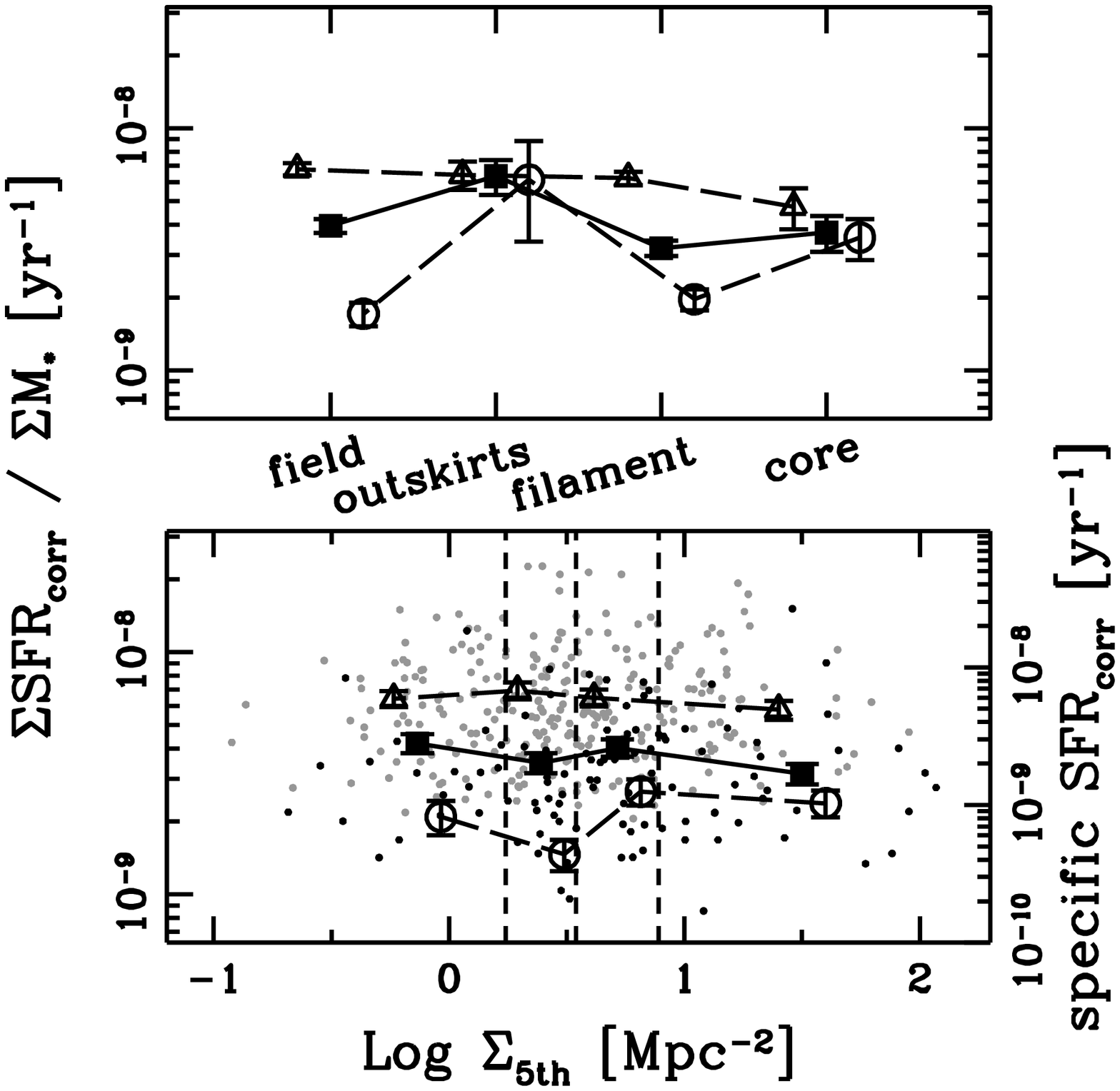}
 \end{center}
 \caption{The integrated SSFR ($\Sigma$SFR/$\Sigma$\Mstar) as a
   function of environment. The integrated SSFR is defined as the integrated
   SFR ($\Sigma$SFR) divided by the integrated stellar mass
   ($\Sigma$\Mstar) for the \oii\ emitters in each region. Filled squares
   show the integrated SSFRs for all the \oii\ emitters. Open circles show
   those of \oii\ emitters with \Mstar$>10^{10.5}$\Msun, while open
   triangles show those of the \oii\ emitters with \Mstar$<10^{10.5}$\Msun.
   The data points are slightly offset manually to avoid their overlapping.
   In the lower panel, black dots show the SSFRs of the individual \oii\ emitters
   with \Mstar$>10^{10.5}$\Msun, and gray dots are those of the \oii\ emitters
   with \Mstar$<10^{10.5}$\Msun.  The right vertical axis shows
   specific SFR.    
}
 \label{fig;ssfr}
\end{figure}

In this section, we derive SFRs and specific SFRs (SSFRs) of the \oii\ emitters 
from their \oii\ line fluxes (ergs\ s$^{-1}$ cm$^{-2}$) and continuum flux
densities (ergs\ s$^{-1}$ cm$^{-2}$ \AA$^{-1}$).
They are calculated from flux densities in NB912 and $z'$ bands ($f_{NB912}$ and
$f_{z'}$), respectively, as follows;  
\begin{equation}
F({\rm [O\,{\scriptstyle II}]})=f_{NB912}\Delta_{NB912}\frac{1-(f_{z'}/f_{NB912})}{1-(\Delta_{NB912}/\Delta_{z'})},
\label{eq;fline}
\end{equation}
\begin{equation}
f_{\lambda,{\rm cont}}=f_{z'}\frac{1-(f_{NB912}/f_{z'})(\Delta_{NB912}/\Delta_{z'})}{1-(\Delta_{NB912}/\Delta_{z'})},
\label{eq;fcont}
\end{equation}
where $\Delta_{NB912}$ and $\Delta_{z'}$ indicate FWHMs of the filters,
and $\Delta_{NB912}=134$\AA\ and $\Delta_{z'}=955$\AA. 

The SFR is the fundamental quantity to characterize the star formation activity
in a galaxy.  We convert a \oii\ luminosity into a SFR using 
the calibration by \citet{kennicutt1998}.
For a dust extinction correction, we take the following empirical approach. 
\citet{garn2010} have found a correlation between the observed \ha\
luminosity and the dust attenuation for \ha\ emitters at $z=0.845$
under the High-$z$ Emission Line Survey (HiZELS)
\citep[equation (6) in][]{garn2010};   
\begin{equation}
A_{\rm H\alpha} = -19.46 + 0.50 \log_{10}\left( \frac{L_{\rm H\alpha,obs}}{\rm erg\ s^{-1}} \right).
\label{eq;garn2010}
\end{equation}
Because the observed luminosity ratio between \oii\ and \ha\ is sensitive to
dust extinction, the assumption of the constant $L_{\rm [OII]}/L_{\rm  H\alpha}$
is inadequate. \citet{moustakas2006} have
found that the correlation between the observed ratio of \oii\ to \ha\
and $E(B-V)$ is consistent with the Galactic extinction curve of
\citet{odonnell1994} for local star-forming galaxies, and we get;
\begin{eqnarray}
\log_{10}\left(L_{\rm [OII]}/L_{\rm H\alpha}\right)_{\rm obs} & = & -0.861\ E(B-V),\\
 & = & -0.342\ A_{\rm H\alpha},
\label{eq;moustakas2006}
\end{eqnarray}
where $A_{\rm H\alpha}=2.52E(B-V)$ assuming the Galactic extinction curve of
\citet{odonnell1994}. Under the assumption that the two relations of
(\ref{eq;garn2010}) and (\ref{eq;moustakas2006}) are valid for the \oii\
emitters at $z=1.46$, we use them to derive a \ha\ luminosity,
$L_{\rm H\alpha,obs}$, and a dust attenuation index, $A_{\rm H\alpha}$, from
the observed \oii\ luminosity. Then, the dust-corrected \ha\
luminosity is converted into a SFR$_{\rm corr}$ using the relation
of \citet{kennicutt1998}.  
It should be noted that the amounts of dust attenuation are comparable to
those derived from the Balmer decrement of \hb/\ha\ (\S~\ref{sec;dust-extinction}).
Furthermore, we find that $E(B-V)$s derived
from broad-band colours in $B-z'$ using the equation (4) of \citet{daddi2004}
are consistent with
those derived from \oii\ luminosities under the assumption that the ratio
between stellar and nebulous components in dust extinction is 0.44
\citep{calzetti2000}, although the scatter is large.
These facts support that our dust correction method is valid.

For the \oii\ emitters at $z=1.46$, $K$-band corresponds to the rest-frame
$z'$-band, i.e.\ $\sim$8900\AA. We estimate stellar masses of the
\oii\ emitters using an empirical relation between stellar mass for
$K$-selected galaxies at $z>1.4$ and its $K$ magnitude given in
\citet{daddi2004}. The mass-to-light ratio of a galaxy is calibrated
with its $z-K$ colour, which can reduce the dispersion of derived
stellar mass to $\sigma(\Delta \log({\rm M_\star}))=0.20$ \citep{daddi2004}.
Using thus derived stellar mass of individual galaxies, we also derive
SSFR by dividing SFR by the stellar mass.   

Fig.~\ref{fig;mass-sfr} shows SFRs of the \oii\ emitters in
each environment as a function of stellar mass. There is a weak
correlation that more massive galaxies tend to have larger SFRs,
except for the \oii\ emitters in the field region.
It is also notable that, although there is a considerable scatter,
for a given stellar mass, the \oii\ emitters in the cluster have
similar SFR$_{\rm corr}$ to those of field galaxies at $z=1.5-2.5$
\citep{erb2006b,daddi2007,hayashi2009,yoshikawa2010}.
This suggests that, at $z=1.46$ the environmental dependence of 
star formation activity is weak, and that galaxies in the XCS2215
cluster at $z=1.46$ have conducted strong star formation activities
just comparable to those in the general field.

Fig.~\ref{fig;mass-ssfr} shows SSFR as a function of stellar mass,
which suggests that more massive galaxies have lower SSFR.
Massive galaxies do not necessarily enhance their star formation
activity. We investigate the global SSFR, which are calculated
from integrated SFR$_{\rm corr}$ divided by integrated stellar mass for
the \oii\ emitters in each environment (Fig.~\ref{fig;ssfr}). This figure
also supports that star formation activity of galaxies at $z=1.46$ is
not dependent on environment strongly.

At redshifts below unity or so, the star formation activity in clusters is
significantly weaker than that in the surrounding regions.
In RXJ1716.8+6708 cluster at
$z=0.81$ and XMMU~J2235.3-2557 cluster at $z=1.39$, no star-forming
galaxies is found within $\sim$200-250kpc from the cluster centre
\citep{koyama2010,lidman2008,rosati2009,bauer2011}.
However, \citet{hayashi2010} and \citet{hilton2010} found that there are
a significant fraction of star-forming galaxies near the centre of the XCS2215 cluster. 
\citet{fassbender2011} also found an on-going starburst activity in
XMMU~J1007.4+1237 cluster at $z=1.56$. Most of the clusters at $z\ga1.5$
are likely to hold active star formation in the core regions.
In order to make an evolutionary link to the lower-redshift clusters
with inactive cores, some processes must take place to suppress
star formation in the central regions in the short time interval
between $z\sim1.5$ and $z\sim1.0$ ($\la$ 1.5~Gyrs).

\section{Cosmic evolution of star formation activity in clusters}
\label{sec;evolution_sf}

In this section, we compare the global SSFR of the XCS2215 cluster with
those of clusters at lower redshifts. For clusters at $z<1$, the
redshift evolution of the global SSFRs approximately follow a relation of
$\sim(1+z)^6$ although the scatter is large \citep[e.g.,][]{finn2005,koyama2010}. 
For comparison with those previous studies, we derive integrated SSFR
and dynamical mass of the XCS2215 cluster in the same manner as in the
previous works. 

To derive the integrated SFR ($\Sigma$SFR) in the XCS2215 cluster,
we sum up individual SFR$_{\rm corr}$ of the \oii\ emitters within
a radius of 0.5$\times R_{200}$, where $R_{200}$ is a radius within which
the averaged matter density is 200 times larger than the critical density.
Previous studies apply a constant dust extinction, \Aha=1, to derive
intrinsic SFRs, which is different from our correction for dust
extinction adopted in this paper. Thus, for this comparison only,
we apply the same amount of correction, \Aha=1.
The cluster mass, $M_{cl}$, is estimated from the
velocity dispersion of the cluster, which is 720 km s$^{-1}$ 
\citep{hilton2010}. 
Then, we use equations (4) and (5) in \citet{koyama2010} to derive
$R_{200}$ and $M_{cl}$ for the XCS2215 cluster. The radius, $R_{200}$, is
0.8 Mpc which corresponds to 1.57 arcmin \citep{hilton2010}. The mass,
$M_{cl}$, is $2.81\times10^{14}$\Msun. As a result, the integrated
SFR, $\Sigma$SFR, is 764$\pm$23.3\Msun yr$^{-1}$.

Before comparing with the previous studies, there are two issues to
keep in mind. One is that $R_{200}$ and $M_{cl}$ derived from the
velocity dispersion may be overestimated. \citet{hilton2010} pointed
out that the redshift distribution of cluster members shows a bimodality,
and thus this cluster may have been experiencing a merger event in the
recent past within a few Gyr or so.
Our spectroscopy also shows a similar redshift distribution for
the \oii\ emitters. 
The other issue is that our \oii\ survey may be underestimating
the integrated SFR compared to the \ha\ surveys. Our SSFRs are
derived from \oii\ luminosities, while the SSFRs in the other
studies are derived from \ha\ luminosities.
Our \oii\ survey is probably more sensitive to dust extinction and
the depth (2.6\Msun yr$^{-1}$) is also slightly shallower than the
\ha\ surveys ($<$1\Msun yr$^{-1}$) at lower redshifts in terms of
dust-free limiting star formation rates.  
In spite of such differences between our study and the previous ones,
however, it is still worth comparing among these results.

\begin{figure}
 \begin{center}
 \includegraphics[width=0.9\linewidth]{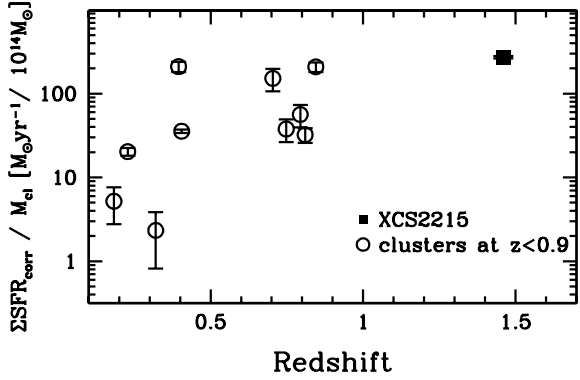}
 \end{center}
 \caption{ The integrated SFR per unit cluster mass
 ($\Sigma$SFR$_{\rm corr}$/$M_{\rm cl}$) is plotted as a function of redshift.
 The integrated SFR, $\Sigma$SFR$_{\rm corr}$, is
 a sum of the SFR corrected for dust extinction assuming $A_{\rm H\alpha}=1$. 
 The cluster mass, $M_{\rm cl}$, is derived from the velocity
 dispersion. The filled square shows our data point of the XCS2215 cluster.
 Open circles show other clusters taken from the literature; 
 A1689 ($z=0.183$) from \citet{balogh2002}, 
 A2390 ($z=0.228$) from \citet{balogh2000}, 
 AC114 ($z=0.320$) from \citet{couch2001},
 CL0024.0+1652 ($z=0.395$) from \citet{kodama2004}, 
 A851 ($z=0.41$) from \citet{koyama2011}, 
 CL1040 ($z=0.704$), CL1054-12 ($z=0.748$) and CL1216 ($z=0.794$) from
 \citet{finn2005}, RXJ1716($z=0.81$) from \citet{koyama2010},
 and CLJ0023+0423B ($z=0.845$) from \citet{finn2004}. 
  }
\label{fig;ssfr_redshift}
\end{figure}

Fig.~\ref{fig;ssfr_redshift} shows the SSFRs of clusters as a
function of redshift. The XCS2215 cluster has the largest SSFR among
the clusters plotted in the figure, and there is a general trend that
the star formation activity in galaxy clusters increases with 
redshift out to $z\sim1.46$.
It should be noted that both of the two issues described above lead to
the conservative estimation of SSFR for the XCS2215, which would
therefore strengthen the trend that we claim.
This result seems reasonable, because the cosmic star formation rate density
keeps rising to $z\sim2$ on average and the redshift of $z=1.46$ is
closer to the epoch when galaxy clusters are formed.
\citet{tadaki2011} suggest that the star formation activity in
cluster/proto-cluster may be even stronger than in the field at $z\sim2$.
As suggested in \S~\ref{sec;phot-properties}, the redshift range of
$z=$1.5--2.5 is probably the epoch when galaxies form stars very actively
irrespective of their environments.

\section{Spectroscopic properties of the \oii\ emitters}
\label{sec;spec-properties}

\begin{table*}
 \caption{Line properties of our 16 spectroscopically confirmed \oii\ emitters. 
  The observed \oii\ luminosities, ratios of emission
  lines, color excesses, and oxygen abundances are shown.
  }
\begin{center}
\begin{tabular}{cccccccccl}
\hline\hline
ID & \oii\ flux$^{\dagger,a}$ & \oii\ luminosity$^{\dagger,b}$ & \ha/\hb & $E(B-V)$$^c$ & \nii/\ha & Z(N2)$^d$ & \oiii/\hb & Z(O3\hb)$^e$  & Comment \\
\hline
 1 & 0.58$\pm$0.08 & 0.77$\pm$0.11 & --- & --- & 0.29$\pm$0.10 & 8.59$^{+0.07}_{-0.11}$ & --- & --- & --- \\ 
 2 & 3.90$\pm$0.14 & 5.15$\pm$0.19 & --- & --- & --- & --- & $>$21.40 & --- & AGN \\ 
 3 & 0.54$\pm$0.13 & 0.73$\pm$0.17 & --- & --- & $<$0.28 & $<$8.582 & --- & --- & --- \\ 
 4 & 1.17$\pm$0.09 & 1.56$\pm$0.12 & --- & --- & --- & --- & 2.80$\pm$1.26 & 8.46$^{+0.21}_{-0.18}$ & --- \\ 
 5 & 0.41$\pm$0.09 & 0.55$\pm$0.12 & --- & --- & 0.34$\pm$0.23 & 8.63$^{+0.13}_{-0.28}$ & --- & --- & --- \\ 
 6 & 0.72$\pm$0.10 & 0.93$\pm$0.13 & --- & --- & $<$0.48 & --- & $>$6.77 & --- & AGN \\ 
 7 & 1.33$\pm$0.13 & 1.75$\pm$0.18 & 5.06$\pm$1.57 & $0.54^{+0.36}_{-0.26}$ & 0.54$\pm$0.20 & 8.75$^{+0.08}_{-0.11}$ & $<$0.63 & $>$8.899 & --- \\ 
 8 & 2.30$\pm$0.13 & 3.06$\pm$0.17 & 7.44$\pm$2.54 & $0.91^{+0.40}_{-0.28}$ & 0.32$\pm$0.08 & 8.62$^{+0.05}_{-0.07}$ & 3.87$\pm$1.35 & 8.31$^{+0.20}_{-0.24}$ & --- \\ 
 9 & 1.64$\pm$0.12 & 2.20$\pm$0.16 & --- & --- & --- & --- & $>$7.94 & --- & AGN \\ 
10 & 1.27$\pm$0.09 & 1.67$\pm$0.12 & 4.36$\pm$0.81 & $0.40^{+0.20}_{-0.16}$ & 0.45$\pm$0.08 & 8.70$^{+0.04}_{-0.05}$ & 0.86$\pm$0.32 & 8.83$^{+0.11}_{-0.08}$ & --- \\ 
11 & 0.98$\pm$0.11 & 1.31$\pm$0.15 & --- & --- & --- & --- & $>$2.19 & $<$8.555 & --- \\ 
12 & 0.94$\pm$0.12 & 1.27$\pm$0.17 & --- & --- & --- & --- & 2.02$\pm$0.58 & 8.58$^{+0.11}_{-0.09}$ & --- \\ 
13 & 0.23$\pm$0.07 & 0.29$\pm$0.09 & --- & --- & --- & --- & $<$0.86 & $>$8.824 & --- \\ 
14 & 0.29$\pm$0.10 & 0.40$\pm$0.13 & --- & --- & --- & --- & $>$3.15 & $<$8.412 & --- \\ 
15 & 1.04$\pm$0.09 & 1.39$\pm$0.12 & 5.97$\pm$1.33 & $0.70^{+0.24}_{-0.19}$ & 0.09$\pm$0.04 & 8.30$^{+0.09}_{-0.15}$ & 5.63$\pm$1.14 & 7.87$^{+0.35}_{-0.35}$ & --- \\ 
16 & 0.38$\pm$0.08 & 0.51$\pm$0.11 & --- & --- & --- & --- & $>$3.10 & $<$8.419 & --- \\ 
\hline\hline
\multicolumn{10}{l}{ $\dagger$ Corrected for the filter response based on the spectroscopic redshift.} \\
\multicolumn{10}{l}{ $a$ The numbers are in the units of 10$^{-16}$ erg s$^{-1}$ cm$^{-2}$. } \\
\multicolumn{10}{l}{ $b$ The numbers are in the units of 10$^{42}$ erg s$^{-1}$. } \\
\multicolumn{10}{l}{ $c$ $E(B-V)$ is derived from the Balmer decrement (\ha/\hb\ ratio).} \\
\multicolumn{10}{l}{ $d$ Z(N2) is oxygen abundance, 12+log(O/H), estimated from the \nii/\ha\ ratio. } \\
\multicolumn{10}{l}{ $e$ Z(O3\hb) is oxygen abundance, 12+log(O/H), estimated from the \oiii/\hb\ ratio.} \\
\label{tbl;line-ratio}
\end{tabular}
\end{center}
\end{table*}

In this section, we discuss spectroscopic properties of the \oii\
emitters using the ratios of detected emission
lines. Table~\ref{tbl;line-ratio} shows the observed \oii\
luminosities and the ratios of emission lines.

\subsection{AGN contribution}
\label{sec;AGN}

\begin{figure}
 \begin{center}
 \includegraphics[width=\linewidth]{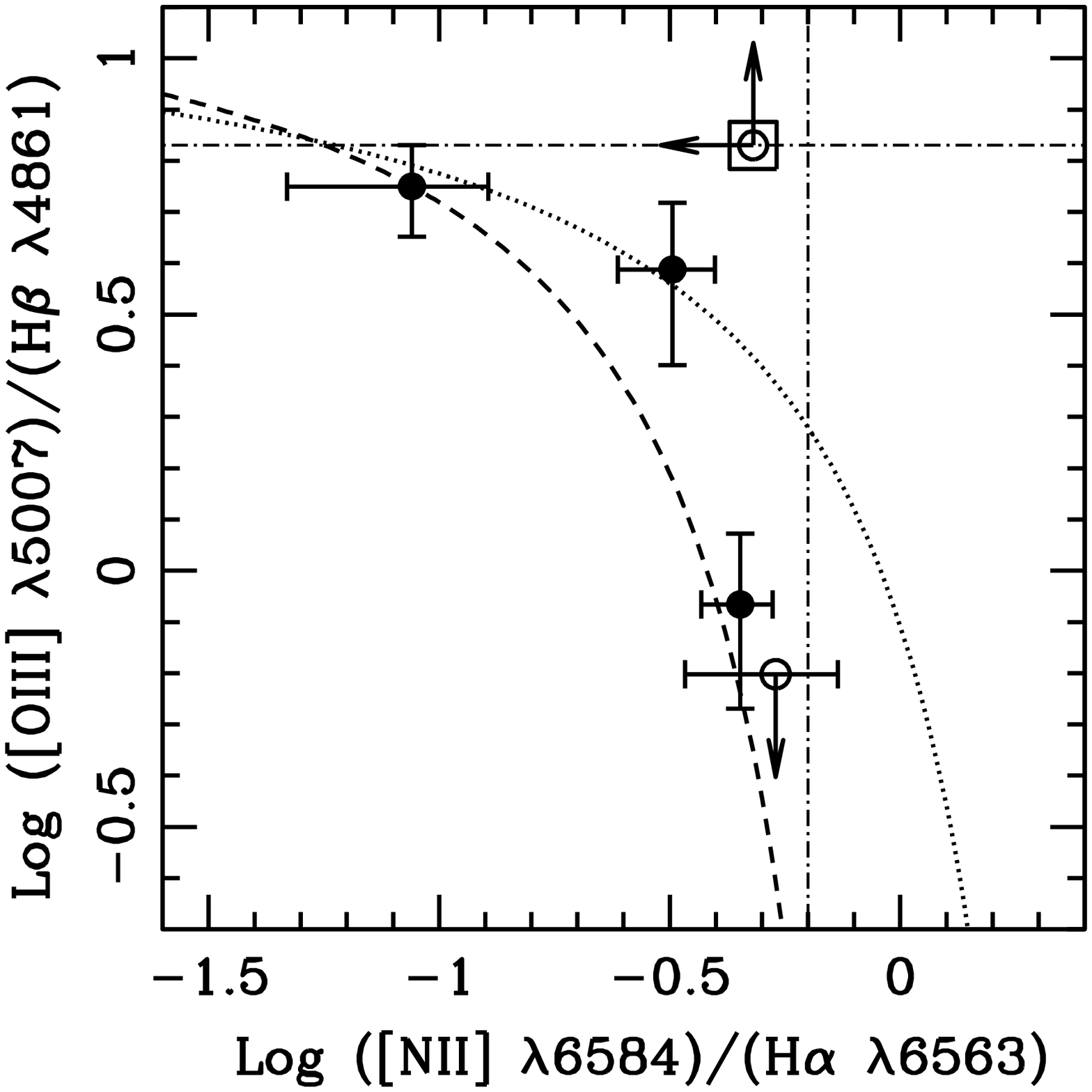}
 \end{center}
 \caption{ The emission-line diagnostic diagram showing 
  \oiii($\lambda$5007)/\hb\ vs. \nii($\lambda$6584)/\ha\ ratios. 
  The dashed curve shows the boundary separating star forming galaxies
  and AGNs based on the SDSS data \citep{kauffmann2003}.
  The dotted curve shows a theoretical boundary given in
  \citet{kewley2001}.
  Star forming galaxies are located on the bottom/left side of the curves,
  while AGNs are located on the top/right side.
  The region between the two curves is a composite region, where both
  star formation and AGN are contributing.
  Filled circles show the \oii\ emitters with all the lines detected,
  while open circles indicate those with some lines un-detected and 
  2$\sigma$ sky noise levels are shown as upper/lower limits.
  The definitive AGN is marked by an open square.  
  The dot-dashed lines show
  Log(\nii($\lambda$6584)/\ha)=-0.2 and
  Log(\oiii($\lambda$5007)/\hb)=0.83, which are used to separate AGNs
  when only either of the ratios can be measured (Fig.~\ref{fig;line_ratio}).
}
 \label{fig;bpt}
\end{figure}

\begin{figure}
 \begin{center}
 \includegraphics[width=\linewidth]{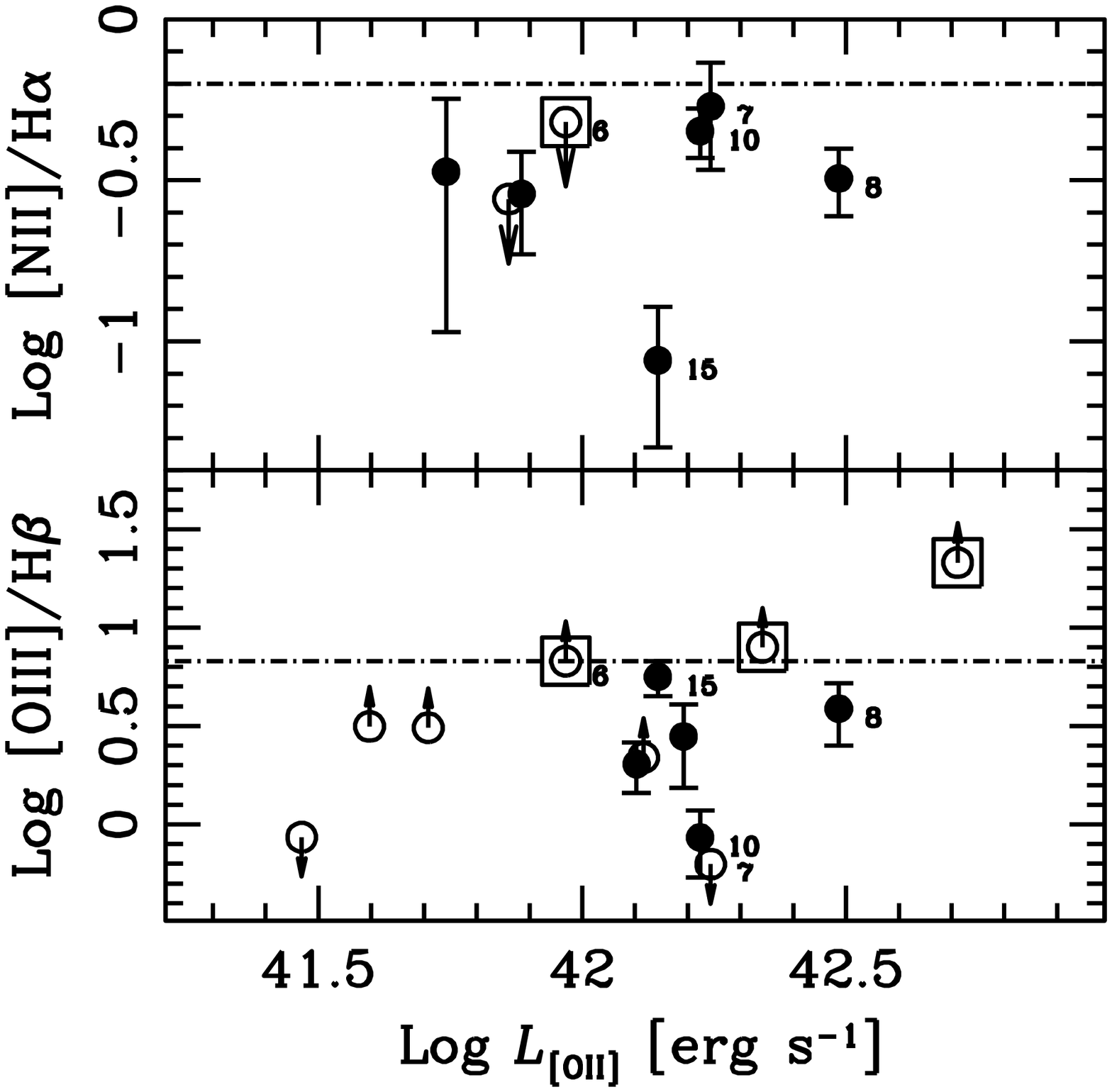}
 \end{center}
 \caption{ The emission-line ratios as a function of \oii\ luminosity. 
  The upper panel shows \nii($\lambda$6584)/\ha\ ratios, while the
  lower panel shows \oiii($\lambda$5007)/\hb\ ratios. Filled circles
  indicate the objects with both of the lines detected. Open
  circles indicate those without either of the lines,
  and thus these line ratios are upper/lower limits.
  The dot-dashed lines show Log(\nii($\lambda$6584)/\ha)=-0.2 and
  Log(\oiii($\lambda$5007)/\hb)=0.83. Three AGNs candidates are marked
  by open squares.
  The numbers indicated beside the symbols correspond to the ID
  numbers of the \oii\ emitters as listed in Table 4 for comparison.   
 }
 \label{fig;line_ratio}
\end{figure}

Since the AGN activity in the Universe peaks at $z\sim2$
\citep[e.g.,][]{ueda2003}, it is important to investigate a
contribution of AGN to our \oii\ emitters before discussing their
spectroscopic properties. \citet{yan2006} investigated the origin of
\oii\ emission lines from local galaxies using the SDSS data, and
concluded that \oii\ emission lines from red galaxies are dominated by
radiation from low-ionization nuclear emission-line regions (LINERS),
rather than star-forming H{\sevensize II} regions. On the other hand,
\oii\ emission lines from blue galaxies mainly come from star
formation. \citet{lemaux2010} conducted NIR spectroscopy of
LINER-type galaxies in clusters at $z$=0.8--0.9 and obtained similar
conclusions to those of \citet{yan2006}.
Although most of our \oii\ emitters have blue colours hence their
\oii\ emissions are likely due to star formation, these studies
demonstrate the importance of evaluating the AGN contribution in our \oii\
emitters. 

The two emission line ratios of \nii/\ha\ and \oiii/\hb\ are
frequently utilized to distinguish star-forming galaxies from
AGNs \citep[e.g.,][]{baldwin1981,kewley2001,kauffmann2003}.
Fig.~\ref{fig;bpt} shows a \nii/\ha\ vs.~\oiii/\hb\ diagram for our
\oii\ emitters in the XCS2215 cluster.  
The dashed line shows a boundary separating between star forming
galaxies and AGNs, which is defined based on the SDSS data \citep{kauffmann2003}.
The dotted line shows a theoretical boundary given by
\citet{kewley2001}. For our spectroscopic sample, there are only three
\oii\ emitters with all the emission lines detected (see
Table~\ref{tbl;spec_sample}). If either of the line ratios is not
available, we evaluate an upper or lower limit
of the ratio by assuming a Gaussian profile with a peak flux that is
twice the sky noise and $\sigma=10$\AA\ for the non-detected line. We
note that the assumed width of the Gaussian profile is comparable to
those of the detected lines.  We show the galaxies with such
upper/lower limit(s) by open circles with arrow(s) in Fig.~\ref{fig;bpt}.   
Fig.~\ref{fig;line_ratio} shows \nii/\ha\ and \oiii/\hb\ ratios, separately.
The number of objects shown in the plots is significantly increased.
The dot-dashed lines indicate Log(\nii/\ha)$=-0.20$ and Log(\oiii/\hb)$=0.83$,
respectively, above which emission line ratios are more AGN-like.

Almost all the \oii\ emitters for which all the four emission lines
are available are located in the composite region between the dotted
and the broken lines in Fig.~\ref{fig;bpt}. Only one of them is
clearly in the AGN region. Fig.~\ref{fig;line_ratio} also shows that
few \oii\ emitters have a large contribution of AGN. Only three \oii\
emitters are located above the threshold lines for AGN. Therefore, it
is unlikely that our \oii\ emitters are dominated by AGNs. Although the
red \oii\ emitters may have more contribution of LINER/Seyfert-type AGNs
as pointed out by the previous studies \citep{yan2006,lemaux2010},
Fig.~\ref{fig;cmd} shows that a large fraction of our \oii\ emitters
are distributed in the blue cloud, and thus most of the \oii\ emission
lines are more likely to be originating from star formation activity
rather than AGN. At the same time, however, AGN contribution is not
negligible, and a part of the \oii\ emission line fluxes may come from
AGNs. 

\citet{hilton2010} detected two X-ray point sources with Chandra
among the spectroscopically confirmed members of XCS2215 cluster,
whose limiting fluxes are 
$\approx1.0\times10^{-16}$ erg s$^{-1}$ cm$^{-2}$ corresponding to 
$L_{\rm X(2-10keV)}\ga0.8\times10^{42}$ erg s$^{-1}$ at $z=1.46$
if the spectral index of $\alpha=2$ and $N_{\rm H}=1.0\times10^{22}$ cm$^{-2}$
are assumed \citep{hilton2010}.
Moreover, they found that one of the 24\micron\ sources has a
mid-infrared SED based on the Spitzer data that is consistent
with an AGN although none of the 24\micron\ sources are detected in the
X-ray data. Among these three AGN candidates, two of them (a X-ray
source and a 24\micron\ source) are included in our \oii\ emitter
sample. The X-ray source is ID-2 of the \oii\ emitters, while no
line is detected in our MOIRCS spectrum for the 24\micron\
source. This suggests that most of our \oii\ emitters are not heavily
contaminated by AGN activities, except for the red \oii\ emitters
(\S~\ref{sec;cmd} and Fig.~\ref{fig;zjk_colour}).   

All these results support that the XCS2215 cluster has active star
forming activities even in the core of the cluster as reported in
\citet{hayashi2010} and \citet{hilton2010}.
As Figs~\ref{fig;bpt} and \ref{fig;line_ratio} indicate,
we cannot completely ignore the contribution from AGNs,
and in principle, we must take care of the influence of AGN on the line
flux when we discuss the properties based on the emission line fluxes.
However, since it is impossible to quantitatively measure
a contribution of AGN to the line flux with the currently available data,
we have to assume at this stage that \oii\ fluxes originate purely from
star formation in \S\ref{sec;sfr_ssfr}.
It should be noted therefore that the SFRs derived from the \oii\
luminosities are probably overestimated.

\subsection{Dust extinction}
\label{sec;dust-extinction}

The galaxy properties derived from the observables (colours and line intensities)
are sensitive to the correction for dust extinction.
In order to estimate the amount of dust extinction for stellar continuum flux,
the SED fitting to the multi-band photometric data is frequently conducted.
However, it is hard to break the degeneracy between stellar age and dust reddening.
Also the amount of dust extinction for the nebular emission lines from
H{\sevensize II} regions is different from that of the stellar
continuum SEDs. A large uncertainty exists in the conversion from
stellar reddening to nebular reddening.

One of the reliable methods to derive the amount of dust extinction of
the nebular emission is to use the Balmer decrement (i.e.~\ha/\hb\ ratio).
Under the assumption of the Case B recombination, intrinsic flux ratio of the two
Balmer lines is expected to (\ha/\hb)$_{\rm int}=2.86$.
By comparing the observed flux ratio, (\ha/\hb)$_{\rm obs}$, with the intrinsic one,
we can estimate the amount of dust extinction as follows;
\begin{equation}
E(B-V)=-\frac{2.5}{k(\lambda_{H\beta})-k(\lambda_{H\alpha})}\log\left\{\frac{(H\alpha/H\beta)_{\rm int}}{(H\alpha/H\beta)_{\rm obs}}\right\},
\end{equation}
where $k(\lambda_{H\beta})-k(\lambda_{H\alpha})=1.14$ as calculated
from the \citet{odonnell1994} extinction curve
(as in \S\ref{sec;sfr_ssfr}).

There are four \oii\ emitters in our spectroscopic sample for which
both of the Balmer lines are detected. The derived amounts of dust extinction are 
$E(B-V)$ = $0.543^{+0.355}_{-0.258}$,
$0.910^{+0.397}_{-0.279}$, $0.401^{+0.197}_{-0.163}$, and
$0.700^{+0.240}_{-0.192}$ for ID-7, ID-8, ID-10 and ID-15,
respectively.
These colour excesses correspond to \Aha= $1.37^{+0.89}_{-0.65}$, 
$2.29^{+1.00}_{-0.70}$, $1.01^{+0.50}_{-0.41}$, and
$1.76^{+0.60}_{-0.48}$, respectively. 
Note that these $E(B-V)$ values are comparable to those of
the $BzK$-selected field galaxies \citep{yoshikawa2010}. 
The estimated dust extinction is larger than the nominal value
of \ha\ flux, \Aha=1 that is frequently used in the literature.
If we correct \oii\ fluxes only by the amount corresponding to \Aha=1 for
all our emitters uniformly, the intrinsic \oii\ fluxes are likely to be
underestimated significantly.
 
For the two \oii\ emitters of ID-6 and ID-8, both \oii\ and \ha\ lines are
detected. The observed ratio of the two lines can provide us with
another estimation of \Aha\ using the equation (\ref{eq;moustakas2006}). 
The observed ratios, $L_{\rm [OII]}/L_{\rm H\alpha}$, for ID-6 and
ID-8 \oii\ emitters are 0.393$\pm$0.217 and 0.438$\pm$0.066, respectively. 
These correspond to \Aha=$1.19^{+1.02}_{-0.56}$ and
$1.05^{+0.21}_{-0.18}$, respectively. For the ID-8, two independent measures
of dust attenuation differ by a large amount ($\sim$1 mag)
although the uncertainties of the both measurements are large.
Such discrepancy may be caused by a contribution of an AGN to the line fluxes.

In \S \ref{sec;phot-properties}, we use the dust extinction law which is
dependent on SFR (i.e., \ha\ luminosity). Here, we can verify whether
the \citet{garn2010} relation is valid for our \oii\ emitters using
the spectroscopic sample. In the same manner as in \S \ref{sec;phot-properties},
we estimate \Aha\ from the \oii\ flux derived from the narrow-band imaging
but is corrected for filter response using the accurate spectroscopic redshift.
We obtain \Aha=1.9--2.1 based on the \citet{garn2010} relation,
which is consistent with the estimated value by the Balmer decrement technique.
Therefore our method of dust extinction correction based on the
\citet{garn2010} relation is confirmed to be valid for the \oii\ emitters
(\S \ref{sec;phot-properties}).   
However, it is possible that \oii\ fluxes are slightly over-corrected
for dust extinction, and thus SFR$_{\rm corr}$ for the \oii\ emitters may be
a little overestimated.

\subsection{Gas-phase metallicity}

\begin{figure*}
 \begin{center}
 \includegraphics[width=0.48\linewidth]{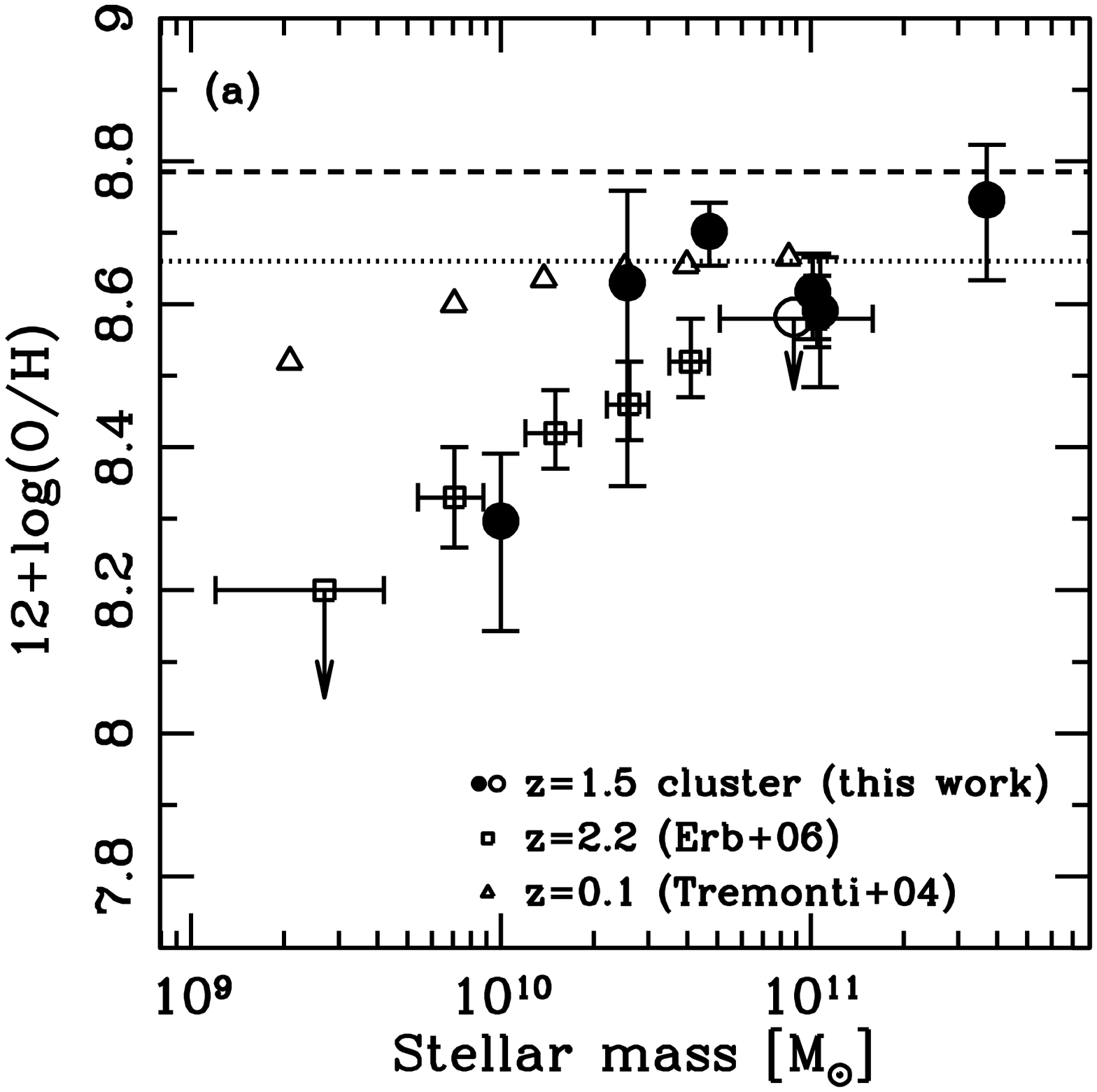}
 \includegraphics[width=0.48\linewidth]{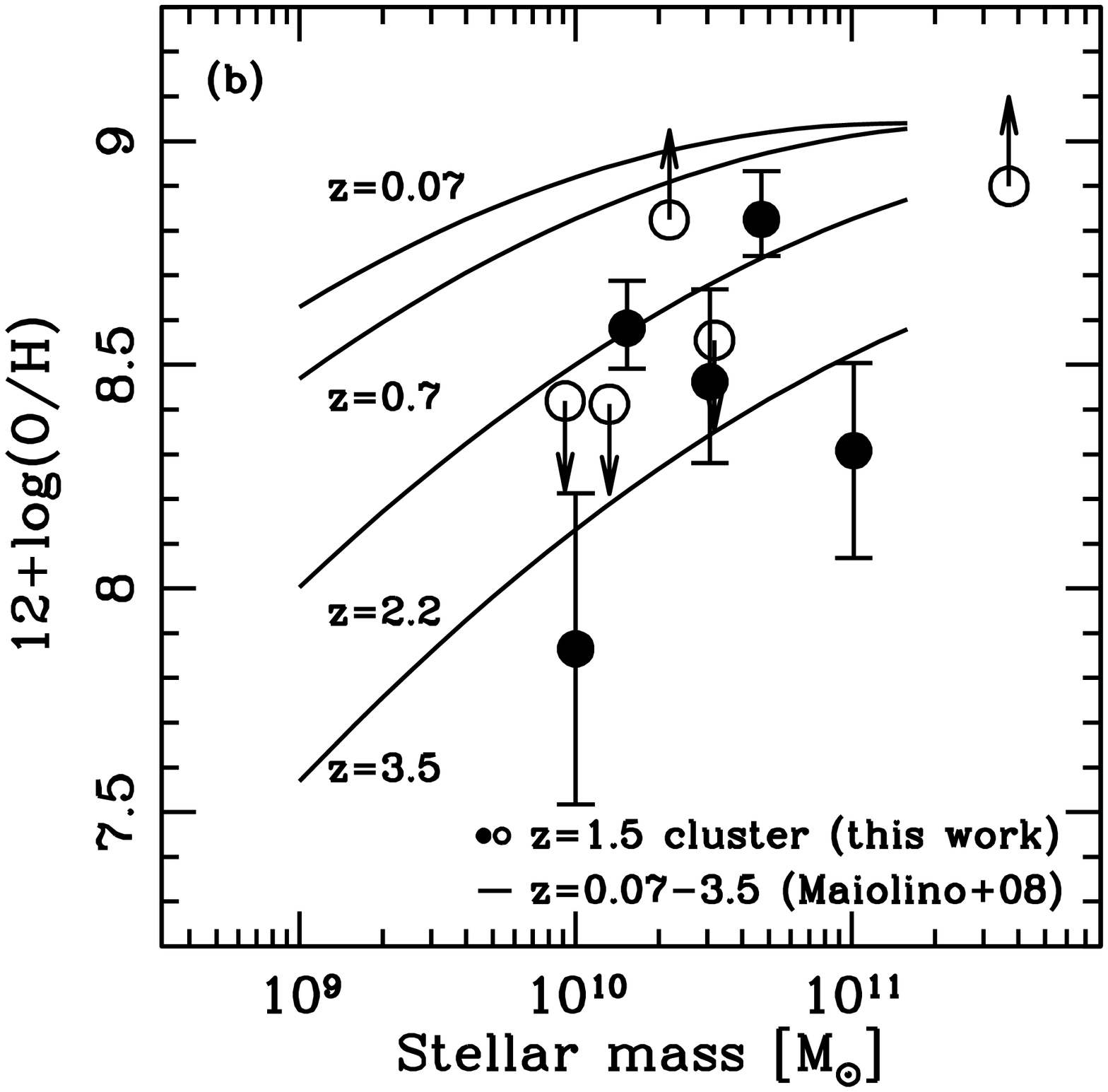}
 \end{center}
 \caption{(a) Left panel: mass--metallicity relation for the \oii\ emitters
   in the XCS2215 cluster. Metallicities are derived from \nii/\ha\ line
   ratios. Filled circles show the objects with both of the lines detected,
   while open circles show those with only either of the lines detected.
   Arrows indicate upper/lower limits in metallicity. 
   Open squares show UV-selected galaxies at $z\sim2$, and open triangles
   indicate the local star-forming galaxies \citep{tremonti2004,erb2006}. The
   dotted line corresponds to the solar oxygen abundance of 12+log(O/H) = 8.66
   \citep{asplund2004}.
   The dashed line shows the abundance for galaxies with the flux
   ratio of log(\nii/\ha)=-0.2. Objects with \nii/\ha\ ratio larger
   than the value are likely to be an AGN (\S\ref{sec;AGN}). 
   (b) Right panel: same as the left panel, but for metallicities
   derived from \oiii/\hb\ line ratios. Solid lines show
   mass--metallicity relations at $z=$0.07, 0.7, 2.2, and 3.5
   \citep{maiolino2008}.
 }
 \label{fig;MZR}
\end{figure*}

It is well known that the electron temperature reflects gas metallicity
\citep{kewley2002,kobulnicky2004,erb2006}. Because the auroral lines
which are used to derive electron temperature are weak and can be
observed only for galaxies with low metallicities,
it is very difficult to estimate gaseous metallicity even for the local
galaxies with this method.
We can only use the ratios of strong emission lines which are emitted from
different ionization levels, to derive gaseous metallicities of high-$z$ galaxies. 
There are several metallicity diagnostics that have been invented,
namely $R_{23}$ and N2 methods which can estimate gas-phase oxygen
abundance, 12+log(O/H).
However, different methods do not always give consistent oxygen abundances.
Even if the same diagnostic is used, different calibrations lead to
very different metallicities \citep{kewley2002,kobulnicky2004}. 
Therefore, one should make sure to use the same diagnostics and calibration
in order to make any proper comparison with other results.

We derive gas phase metallicities of our \oii\ emitters using the two
diagnostics. One is based on \ha/\nii\ flux ratios and its calibration by
\citet{pettini2004}, and the other is based on \hb/\oiii\ flux ratios
and its calibration by \citet{maiolino2008}. 
Both diagnostics are not sensitive to dust extinction, because the
wavelength difference of the pair lines is very small.
If only either of the lines is detected, the upper/lower limit of
metallicity is estimated.  Fig.~\ref{fig;MZR} shows thus derived
metallicities of our \oii\ emitters in the XCS2215 cluster plotted
as a function of stellar mass.
Fig.~\ref{fig;MZR}(a) shows the metallicities from \ha/\nii\ ratios,
while Fig.~\ref{fig;MZR}(b) shows the metallicities from \hb/\oiii\ ratios.
Stellar masses are already derived in \S\ref{sec;phot-properties}.
In both panels, we also show the mass--metallicity relations at different redshifts
taken from the literature \citep{tremonti2004,erb2006,maiolino2008} for comparison.

Although the error bars are large, Fig.~\ref{fig;MZR} suggests that
there is a weak mass--metallicity relation in the sense that more massive
galaxies have higher metallicities.
The metallicities of our \oii\ emitters are comparable to those of
the galaxies at $z$=2--3 \citep{erb2006,maiolino2008}.
Our sample has the galaxies in a high density galaxy cluster, while
the samples of the above previous studies are the galaxies in the
general field at $z$=2--3.
It seems that the star-forming galaxies in the cluster at $z=1.46$ have
a similar mass--metallicity relation to that of the field galaxies at
$z$=2--3. Although the redshift range is slightly different, this may suggest
that the mass--metallicity relation is not very dependent on environment
at $z\ga$1.5.
Here, we must take into account the contribution of AGN to the line flux
ratios. A larger contribution of AGN would make both \nii/\ha\ and \oiii/\hb\ larger,
meaning that the derived metallicities are over-estimated with \nii/\ha\ and
under-estimated with \oiii/\hb, respectively.
As discussed in the previous section, our \oii\ emitters may have a moderate
level of AGN contribution.
In fact, the metallicities estimated by the two different line ratios are
not in good agreement.
However, we can still trust the relative trend such as the existence of
the mass-metallicity relation (but not its absolute values)
as far as the same method is used uniformly.
In order to quantify the evolution of the mass--metallicity relation
at $z\ga1.5$, we need to evaluate the AGN contribution,
but it is beyond the scope of this paper.

The fact that the mass--metallicity relation for the XCS2215 cluster is similar
to that of the field at $z\sim2$, may further support the lack of environmental
dependence, due probably to the high star formation activity in the cluster
cores at this high redshift comparable to that in the field.

\section{Summary}
\label{sec;conclusions}

We conduct a wide-field survey of \oii\ emission line galaxies in and
around the XMMXCS J2215.9-1738 cluster at $z=1.46$ with
Subaru/Suprime-Cam. This survey is an extension of our previous study
reported in \citet{hayashi2010} which was limited to the central
6\arcmin$\times$6\arcmin\ region of the cluster. By combining the UKIRT
$K$ imaging data, we have now extended the analyses to the entire
32\arcmin$\times$23\arcmin\ region.
We investigate colours and star formation activities of the \oii\ emitters
in various environments from the cluster core to the surrounding field at $z=1.46$.   
Moreover, we conduct a follow-up near-infrared spectroscopy of 34 \oii\ emitters
in the cluster core identified in \citet{hayashi2010} with Subaru/MOIRCS,
and obtain $zJ$ spectra covering a wavelength range of
0.9--1.8\micron. These spectra enable us to confirm the existence of \oii\ 
emission lines, and to investigate the detailed properties of the \oii\ emitters
on the basis of multiple nebular emission lines such as \hb, \oiii, \ha, and \nii. 
Our findings are summarized below.

\begin{enumerate}

\item We select 380 \oii\ emitters at $z=1.46$ down to a line flux of 
  $1.4\times10^{-17}$ erg s$^{-1}$ cm$^{-2}$ in our entire survey area of
  32\arcmin$\times$23\arcmin. The spatial distribution of the \oii\
  emitters shows a well-defined filamentary structure of star forming galaxies
  to the east/south of the cluster as well as a concentration of
  such galaxies in the cluster core. The filament is one of the largest
  structures of star-forming galaxies at high redshifts ($1.3\la z\la 3$). 
  Based on the 2D structures, we define four different environments,
  namely, cluster core, outskirts, filament, and the field, in order
  to investigate the environmental dependence of the properties of
  star-forming galaxies at $z=1.46$. 

\item 
  We spectroscopically confirm that at least 16 \oii\ emitters are
  certainly located at $z=1.46$ out of 34 targeted \oii\ emitter
  candidates in the cluster core.  
  Only a single emission line is detected at $\sim$9100\AA\ for three
  candidates. The other 15 candidates have no detected emission 
  lines. However, the \oii\ fluxes estimated by the narrow-band
  imaging suggest that it is likely that most of them have fluxes of
  emission lines smaller than the limiting flux.
  This thus assures that our photometric selection technique is valid
  to efficiently sample \oii\ emitters at $z=1.46$. The redshift
  distribution of the confirmed emitters is consistent with that of
  the confirmed cluster members reported in \citet{hilton2010}, and is
  likely to have a double-peak feature.  

\item The $z'-K$ vs.~$K$ colour--magnitude diagram shows a higher
  fraction of \oii\ emitters on the red sequence in the 
  cluster region than that in the other environments. This suggests
  that some processes which work in the cluster core are responsible for
  the red colours of some \oii\ emitters.
  It is also noted that most of the red \oii\ emitters are massive galaxies. 
  Their SEDs indicate that they are more likely passive galaxies with
  AGNs. We argue therefore that AGN feedback may be a good candidate
  for physical processes to quench star formation activities in
  massive galaxies in high-density regions.

\item The cluster (XMMXCS J2215.9-1738) has been conducting star
  formation at rates comparable to those in other environments.
  This supports our previous study \citep{hayashi2010} which found a high star
  formation activities in the cluster central region.
  It is also found that the global specific star formation rate of
  galaxy cluster, which is calculated as the integrated SFRs divided by
  the integrated stellar masses of galaxies within a radius of
  0.5$\times R_{200}$, increases with redshift up to $z=1.46$.

\item We investigate the flux ratios of emission lines, \oiii/\hb\ and
  \nii/\ha, to distinguish between star-forming galaxies and AGNs.
  On a diagram showing both line ratios, our \oii\ emitters are preferentially
  located in the intermediate, composite region where both star-formation and AGN
  are likely to be contributing.
  It is therefore unlikely that many of the \oii\ emitters are heavily contaminated
  by strong AGNs, but at the same time it is likely that the AGN contribution
  is not negligible.

\item We estimate the strength of dust extinction from the Balmer decrement
  measurements (\hb/\ha) for four of the \oii\ emitters where both lines are
  detected.
  It is found that the extinction index $E(B-V)$ ranges from 0.40 to 0.91,
  suggesting that \oii\ emission line flux is considerably subject to dust
  extinction.

\item We derive gas-phase metallicities for the \oii\ emitters from
  \nii/\ha\ or \oiii/\hb\ line ratios where available.
  It is found that our emitters in the cluster at $z=1.46$ are located
  on a mass--metallicity relation, which is similar to that of the
  star-forming galaxies in the field at $z\sim2$.
  This may also suggest that the star formation activity at
  $z\sim1.5$ and beyond
  is not strongly dependent on environment.

\end{enumerate}

In summary, we find that star formation activity of galaxies at $z=1.46$
is not yet strongly dependent on environment, and that even the
cluster core is experiencing high star forming activity comparable to
those in other lower-density regions.
Our results also suggest that more
detailed understanding of AGN activities along with star formation
activities is crucial to reveal galaxy evolution in clusters in
particular the physical mechanisms of quenching star
formation. Larger, systematic near-infrared spectroscopic surveys such
as those capable with Subaru/FMOS will enable us to understand the
inter-relationship between galaxy and AGN activities, and their
co-evolution. 

\section*{Acknowledgments}
Most of the imaging data and all of the spectroscopic data used in 
this paper are collected at Subaru Telescope, which is operated by the
National Astronomical Observatory of Japan. We thank the Subaru
Telescope staff for their invaluable help to assist our observations
with Suprime-Cam and MOIRCS. A part of NIR data are collected at the
UKIRT, which is operated by the Joint Astronomy Centre on behalf of
the Science and Technology Facilities Council of the U.K. We also
thank the UKIRT staff for their invaluable help to assist our
observations with WFCAM.
We thank Dr. Kentaro Motohara for kindly providing us with his data
reduction package for UKIRT/WFCAM. We would like to thank an anonymous
referee for carefully reading our manuscript and giving useful
comments. 
Y.K.\ acknowledges support from the Japan Society for the Promotion of
Science (JSPS) through JSPS Research Fellowship for Yong Scientists.   
T.K.\ acknowledges support in part by a Grant-in-Aid for the
Scientific Research (No.\ 21340045) by the Japanese Ministry of
Education, Culture, Sports, Science and Technology.

\bsp

\label{lastpage}

\end{document}